\documentclass[10pt, journal, twocolumn]{IEEEtran}
\usepackage[utf8x]{inputenc}

\usepackage[switch, pagewise, modulo, mathlines]{lineno}

\usepackage[numbers, super, sort&compress]{natbib}
\usepackage[makestderr]{pythontex}
\usepackage{multicol}

\usepackage{newtxmath}

\let\labelindent\relax
\let\transp\relax

\usepackage{amsfonts}
\usepackage{mathtools, nccmath}
\usepackage{amsmath} 
\usepackage{amssymb}
\usepackage{bm}

\usepackage[super]{nth}

\usepackage[superscript, noadjust]{cite}

\usepackage{graphicx}
\graphicspath{{figures/}}
\DeclareGraphicsExtensions{.pdf,.jpeg,.png,.eps}

\usepackage{caption}
\usepackage{booktabs}
\usepackage{array}
\usepackage{footnote}
\usepackage{multicol}
\usepackage{color}
\usepackage{enumitem}
\usepackage{stfloats}
\usepackage[hidelinks]{hyperref}
\usepackage[protrusion, expansion, kerning, spacing]{microtype}

\usepackage{siunitx}
\usepackage{xcolor}

\newcommand{\beginsupplement}{%
        \newpage
        \setcounter{table}{0}
        \renewcommand{\thetable}{S\arabic{table}}%
        \setcounter{figure}{0}
        \renewcommand{\thefigure}{S\arabic{figure}}%
     }

\def\trans{{\mathsf{T}}}
\def\herm{{\sfH}}

\def\keywordname{{\emph Keywords:}}%
\def\keywords#1{\par\addvspace\medskipamount{\rightskip=0pt plus1cm
\def\and{\ifhmode\unskip\nobreak\fi\ $\cdot$
}\noindent\keywordname\enspace\ignorespaces#1\par}}

\def\mindex#1{\index{#1}}

\def\sq{\hbox{\rlap{$\sqcap$}$\sqcup$}}
\def\qed{\ifmmode\sq\else{\unskip\nobreak\hfil
\penalty50\hskip1em\null\nobreak\hfil\sq
\parfillskip=0pt\finalhyphendemerits=0\endgraf}\fi\medskip}

\def\sign{{\rm sign}}

\newsavebox{\junk}
\savebox{\junk}[1.6mm]{\hbox{$|\!|\!|$}}

\def\argmax{\mathop{\rm arg\, max}}

\def\bC{{\mathbb C}}

\def\bP{{\mathbb P}}

\def\bR{{\mathbb R}}

\def\bfA{{\bf A}}
\def\bfB{{\bf B}}
\def\bfC{{\bf C}}

\def\bfF{{\bf F}}

\def\bfP{{\bf P}}

\def\bfW{{\bf W}}
\def\bfX{{\bf X}}

\def\bfp{{\bf p}}

\def\bfr{{\bf r}}
\def\bfs{{\bf s}}

\def\bfu{{\bf u}}
\def\bfv{{\bf v}}
\def\bfw{{\bf w}}

\def\sfH{{\sf H}}

\def\sfw{{\sf w}}

\def\bfmath#1{{\mathchoice{\mbox{\boldmath$#1$}}%
{\mbox{\boldmath$#1$}}%
{\mbox{\boldmath$\scriptstyle#1$}}%
{\mbox{\boldmath$\scriptscriptstyle#1$}}}}

\def\bfmY{\bfmath{Y}}

\def\bfmhhaY{\bfmath{\hhaY}} 
\def\bfmhhaY{\hbox to 0pt{$\widehat{\bfmY}$\hss}\widehat{\phantom{\raise 1.25pt\hbox{$\bfmY$}}}}

\def\til={{\widetilde =}}

\def\clG{{\cal G}}

\def\clW{{\cal W}}

 \def\FRAC#1#2#3{\genfrac{}{}{}{#1}{#2}{#3}}

\def\ddtp{{\mathchoice{\FRAC{1}{d^{\hbox to 2pt{\rm\tiny +\hss}}}{dt}}%
{\FRAC{1}{d^{\hbox to 2pt{\rm\tiny +\hss}}}{dt}}%
{\FRAC{3}{d^{\hbox to 2pt{\rm\tiny +\hss}}}{dt}}%
{\FRAC{3}{d^{\hbox to 2pt{\rm\tiny +\hss}}}{dt}}}}

\def\average#1,#2,{{1\over #2} \sum_{#1}^{#2}}

\def\eye(#1){{\bf(#1)}\quad}

\def\eq#1/{(\ref{e:#1})}

\newcommand{\inp}[2]{{\langle #1, #2 \rangle}}

\newcommand{\beqn}[1]{\notes{#1}%
\begin{eqnarray} \elabel{#1}}

\newcommand{\eeqn}{\end{eqnarray} }

\newcommand{\beq}[1]{\notes{#1}%
\begin{equation}\elabel{#1}}

\newcommand{\eeq}{\end{equation}}

\def\bdes{\begin{description}}
\def\edes{\end{description}}

\newcounter{rmnum}

\newcounter{anum}

{\end{list}}

\def\ass(#1:#2){(#1\ref{#1:#2})}

\def\ritem#1{
\item[{\sf \ass(\current_model:#1)}]
}

\newenvironment{recall-ass}[1]{%
\begin{description}
\def\current_model{#1}}{
\end{description}
}

\long\def\comment#1{}

\newfont{\bbb}{msbm10 scaled 700}

\newfont{\bb}{msbm10 scaled 1100}

\renewcommand{\Im}{{\rm Im}}

\newcommand{\transp}{{\sf T}}

\newcommand{\kerD}{{T_\textsc{stht}}}

\def\Rem{{Rem.\,}}

\usepackage{amsthm}
\theoremstyle{definition}
\newtheorem{lemma}{Lemma}
\newtheorem{remark}{Remark}
\newtheorem{theorem}[lemma]{Theorem}
\newtheorem{example}{Example}

\usepackage{geometry}
\renewcommand{\baselinestretch}{1.1}
\usepackage[colorinlistoftodos]{todonotes}

\title{Low-power SNN-based audio source localisation using a Hilbert Transform spike encoding scheme}

\author{Saeid Haghighatshoar$^{1}$, Dylan R. Muir$^{1,*}$ %
    \thanks{%
    $^1$SynSense, Z{\"u}rich, Switzerland.
    This work was partially funded by the ECSEL Joint Undertaking (JU) under grant agreement number 876925, ``ANDANTE''.
    The JU receives support from the European Union's Horizon 2020 research and innovation program and France, Belgium, Germany, Netherlands, Portugal, Spain and Switzerland.
    This work was partially funded by the KDT JU under grant agreement number 101097300, ``EdgeAI''.
    The authors gratefully acknowledge discussions at the Telluride Neuromorphic workshop.
    $^*$Correspondence to Dylan R. Muir, \href{mailto:dylan.muir@synsense.ai}{dylan.muir@synsense.ai}.
    }%
}
\date{October 2023}

\begin{document}

\maketitle
\begin{abstract}
    Sound source localisation is used in many consumer devices, to isolate audio from individual speakers and reject noise.
	Localization is frequently accomplished by ``beamforming'', which combines phase-shifted audio streams to increase power from chosen source directions, under a known microphone array geometry.
	Dense band-pass filters are often needed to obtain narrowband signal components from wideband audio.
    These approaches achieve high accuracy, but narrowband beamforming is computationally demanding, and not ideal for low-power IoT devices.
	We demonstrate a novel method for sound source localisation on arbitrary microphone arrays, designed for efficient implementation in ultra-low-power spiking neural networks (SNNs).
	We use a Hilbert transform to avoid dense band-pass filters, and introduce a new event-based encoding method that captures the phase of the complex analytic signal.
	Our approach achieves state-of-the-art accuracy for SNN methods, comparable with traditional non-SNN super-resolution beamforming.
	We deploy our method to low-power SNN inference hardware, with much lower power consumption than super-resolution methods.
    We demonstrate that signal processing approaches co-designed with spiking neural network implementations can achieve much improved power efficiency.
    Our new Hilbert-transform-based method for beamforming can also improve the efficiency of traditional DSP-based signal processing.
 
\end{abstract}

\section*{Introduction}
Identifying the location of sources from their received signal in an array consisting of several sensors is an important problem in signal processing which arises in many applications such as target detection in radar\cite{skolnik2008radar}, user tracking in wireless systems\cite{haghighatshoar2018low}, indoor presence detection\cite{li2019making}, virtual reality, consumer audio, etc.
Localization is a well-known classical problem and has been widely studied in the literature.

A commonly-used method to estimate the location or the \textit{direction of the arrival} (DoA) of the source from the received signal in the array is to apply reverse beamforming to the incident signals.
Reverse beamforming combines the received array signals in the time or frequency domains according to a signal propagation model, to ``steer'' the array towards a putative target.
The true DoA of an audio source can be estimated by finding the input direction which corresponds to the highest received power in the steered microphone array.
Super-resolution methods for DoA estimation such as MUSIC\cite{schmidt1986multiple} and ESPRIT \cite{roy1989esprit} are among the state-of-the-art methods that adopt reverse beamforming.
Besides source localization, beamforming in its various forms appears as the first stage of spatial signal processing in applications such as audio source separation in the cocktail party problem\cite{haykin2005cocktail, mcdermott2009cocktail} and spatial user grouping in wireless communication\cite{nam2014joint}.

Conventional beamforming approaches assume that incident signals are far-field narrowband sinusoids, and use knowledge of the microphone array geometry to specify phase shifts between the several microphone inputs and effectively steer the array towards a particular direction.\cite{van1988beamforming}
Obviously, most audio signals are not narrowband, with potentially unknown spectral characteristics, meaning that phase shifts cannot be analytically derived.
The conventional solution is to decompose incoming signals into narrowband components via a dense filterbank or Fourier transform approach, and then apply narrowband beamforming separately in each band.
The accuracy of these conventional approaches relies on a large number of frequency bands, which increases the implementation complexity and resource requirements proportionally.

Auditory source localization forms a crucial part of biological signal processing for vertebrates, and plays a vital role in an animal's perception of 3D space\cite{thompson1882li, strutt1907our}.
Neuroscientific studies indicate that the auditory perception of space occurs through inter-aural time- and level-differences, with an angular resolution depending on the wavelength of the incoming signal\cite{yin1990interaural}.

Past literature for artificial sound source localization implemented with spiking neural networks (SNNs) mainly focuses on the biological origins and proof of feasibility of localization based on inter-aural time differences, and can only yield moderate precision in direction-of-arrival (DoA) estimates\cite{wall2012spiking, escudero_real-time_2018, tanoni2019spiking, schoepe2023closed}.
These methods can be seen as array processing techniques based on only two microphones and achieve only moderate precision in practical noisy scenarios.
In this paper, we will not deal directly with the biological origins of auditory localization, but will design an efficient localization method for large microphone arrays (more than two sensors), based on Spiking Neural Networks (SNNs).

SNNs are a class of artificial neural networks whose neurons communicate via sparse binary (0--1 valued) signals known as spikes. 
While artificial neural networks have achieved state-of-the-art performance on various tasks, such as in natural language processing and computer vision, they are usually large, complex, and consume a lot of energy.
SNNs, in contrast, are inspired by biological neural mechanisms \cite{maass1997networks, roy2019towards, panda2020toward, cao2015spiking, dold2022neuro, diehl2015unsupervised} and are shown to yield increases in energy efficiency of several orders of magnitude over ANNs when run on emerging neuromorphic hardware \cite{akopyan2015truenorth, davies2018loihi, moraitis2020optimality, bos2024micro}.

In this work we present a new approach for beamforming and DoA estimation with low implementation complexity and low resource requirements, using the sparsity and energy efficiency of SNNs to achieve an extremely low power profile.
We first show that by using the Hilbert transformation and the complex analytic signal, we can obtain a robust phase signal from wideband audio.
We use this result to derive a new unified approach for beamforming, with equivalent good performance on both narrowband and wideband signals.
We show that beamforming matrices can be easily designed based on the singular value decomposition (SVD) of the covariance of the analytic signal.

We then present an approach for estimating the analytic signal continuously in an audio stream, and demonstrate a new spike-encoding scheme for audio signals that accurately captures the real and quadrature components of the complex analytic signal.
We implement our approach for beamforming and DoA estimation in an SNN, and show that it has good performance and high resolution under noisy wideband signals and noisy speech.
By deploying our method to the SNN inference device Xylo \cite{bos2022sub}, we estimate the power requirements of our approach.
Finally, we compare our method against state of the art approaches for conventional beamforming, as well as DoA estimation with SNNs, in terms of accuracy, computational resource requirements and power.


\section*{Results} 
\subsection*{DoA estimation for far-field audio}
\begin{figure*}[h!] 
    \centering
    \includegraphics[width=120mm]{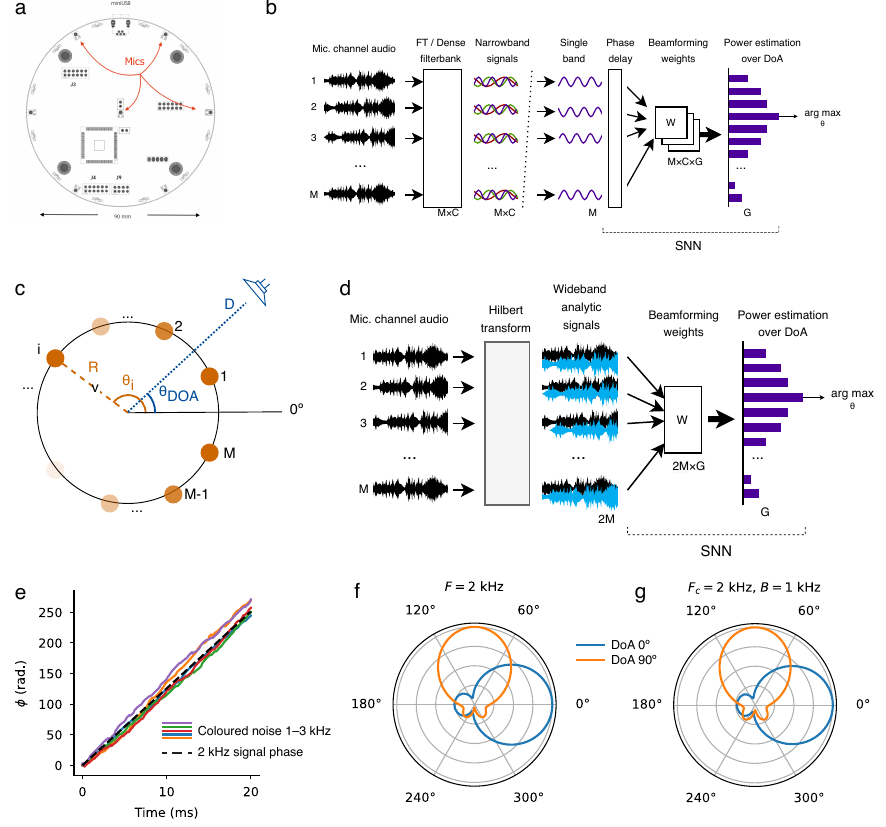}
    \caption{
        \textbf{Narrowband and Hilbert beamforming.}
        \textbf{a} Geometry of a circular microphone array.
        \textbf{b} Narrow-band beamforming approach. A dense filterbank or Fourier transform (FFT/DFT) provides narrowband signals, which are delayed and then combined through a large beamforming weight tensor to estimate direction of incident audio (DoA) as the peak power direction. An SNN may be used for performing beamforming and estimating DoA\cite{pan2021multi}.
        \textbf{c} DoA estimation geometry for far-field audio.
        \textbf{d} Our novel wide-band Hilbert beamforming approach. Wideband analytic signals $X_A$ are obtained by a Hilbert transform. Wideband analytic signals are combined through a small beamforming weight matrix to estimate DoA.
        \textbf{e} The phase progression $\phi$ (coloured lines) of wideband analytic signals $X_A$ generated from wideband noise with central frequency $F_C=2$\,kHz are very similar to the narrowband signal with frequency $F=2$\,kHz.
        \textbf{f--g} Beam patterns from applying Hilbert beamforming to narrowband signals with $F=2$ kHz (f), and wideband signals with center frequency $F_C=2$\,kHz.
    }
    \label{fig:hilbert_beamforming}
\end{figure*}

We examined the task of estimating direction of arrival (DoA) of a point audio source at a microphone array.
Circular microphone arrays are common in consumer home audio devices such as smart speakers.
In this work we assumed a circular array geometry of radius $R$, with $M>2$ microphones arranged over the full angular range $\theta_i \in \left[-\pi, \pi \right]$ (e.g. Figure~\ref{fig:hilbert_beamforming}a).

We assumed a far-field audio scenario, where the signal received from an audio source is approximated by a planar wave with a well-defined DoA.
Briefly, a signal $a(t)$ transmitted by an audio source is received by microphone $i$ as
$x_i(t) = \alpha a\left(t-\tau_i\left(\theta\right)\right) + z_i(t)$,
with an attenuation factor $\alpha$ common over all microphones;
a delay $\tau_i\left(\theta\right)$ depending on the DoA $\theta$;
and with additive independent white noise $z_i(t)$.
For a circular array in the far-field scenario, delays are given by
$\tau_i\left(\theta\right) = {D}/{c} - {R \cos\left(\theta - \theta_i\right)}/{c}$, 
for an audio source at distance $D$ from the array;
a circular array of radius $R$ and with microphone $i$ at angle $\theta_i$ around the array;
and with the speed of sound in air $c=\SI{340}{\meter\per\second}$.
These signals are composed into the vector signal $\textbf{X}(t) = \left(x_1(t), \dots, x_M(t)\right)^\transp$.

DoA estimation is frequently examined in the narrowband case, where incident signals are well approximated by sinusoids with a constant phase shift dependent on DoA, such that $x_i(t) \approx A \sin(2\pi f_0 t - 2\pi f_0 \tau_i(\theta))$.
In this scenario, the signals $\textbf{X}(t)$ can be combined with a set of beamforming weights $\textbf{w}_{\tilde\theta} = (w_{1,\theta}, \dots, w_{M,\theta})^\transp$ to steer the array to a chosen test DoA $\tilde\theta$.
The power in the resulting signal after beamforming $x_b(t; {\tilde\theta}) = \textbf{w}_{\tilde\theta}^\transp \textbf{X}(t)$ given by $P_{\tilde\theta} = \int  |x_b(t; \tilde\theta)  |^2 \textrm{d}t$ is used as an estimate of received power from the direction $\tilde\theta$ and is adopted as an estimator of $\theta$, as the power should be maximal when $\theta = \tilde\theta$.

In practice, the source signal is not sinusoidal, in which case the most common approach is to apply a dense filterbank or Fourier transformation to obtain narrowband components (Figure~\ref{fig:hilbert_beamforming}b).
Collections of narrowband components are then combined with their corresponding set of beamforming weights, and their power is aggregated across the whole collection to estimate the DoA.

\subsection*{Phase behaviour of wideband analytic signals}
\begin{figure}
	\includegraphics[width=\linewidth]{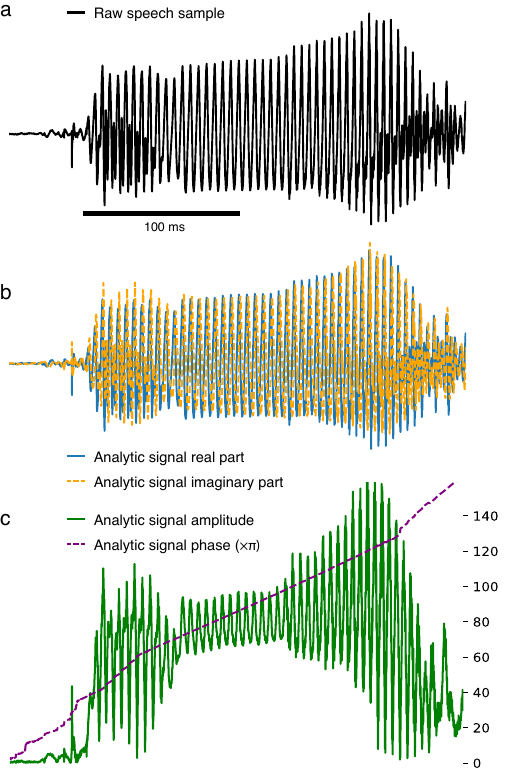}
	\caption{
		\textbf{The Hilbert transform applied to wideband speech signals.}
		\textbf{a} A Raw speech sample.
		\textbf{b} The real (blue) and imaginary (orange) components of the Hilbert analytic signal version of the speech sample.
		Note their similarity to the raw sample, with the phase shift of the imaginary component.
		\textbf{c} The amplitude (green) and unwrapped phase (purple) of the analytic signal.
		Note the smooth phase procession of the analytic signal.
		The instantaneous dominant frequency can be obtained by the slope of the phase signal.
	}
	\label{fig:supp_hilbert_signals}
\end{figure}

Non-stationary wideband signals such as speech do not have a well-defined phase, and so cannot be obviously combined with beamforming weights to estimate DoA.
We examined whether we can obtain a phase counterpart for arbitrary wideband signals, similar to the narrowband case, by applying the Hilbert transform (Figure~\ref{fig:hilbert_beamforming}d; Figure~\ref{fig:supp_hilbert_signals}).

The Hilbert transform is a linear time-invariant operation on a continuous signal $x(t)$ which imparts a phase shift of $\pi/2$ for each frequency component, giving $\hat{x}(t)$.
This is combined with the original signal to produce the \textit{analytic signal} $x_a(t) = x(t) + \textrm{j} \hat{x}(t)$, where `$\textrm{j}$' indicates the imaginary unit.
For a sinusoid $x(t) = \cos(2\pi f_0 t)$, we obtain the analytic signal $x_a(t) = \cos(2\pi f_0 t ) + \textrm{j} \sin(2\pi f_0 t) = \exp(\textrm{j} 2 \pi f_0 t)$.
We can write the analytic signal as $x_a(t) = e(t) \exp(\textrm{j}\phi(t))$, by defining the envelope function~$e(t)$ and the phase function~$\phi(t)$.

If the envelope of a signal $e(t)$ is roughly constant over a time interval $t \in [0, T]$, then $\phi(t)$ is an almost monotone function of $t$, with $\phi(T) - \phi(0) = T \bar{f}$, where $\bar{f}$ is the spectral average frequency of the signal given by

\begin{equation*}
	\bar{f} = \frac{\int_0^\infty {f|X(f)|^2} \textrm{d}f}{\int_0^\infty {|X(f)|^2} \textrm{d}f}	
\end{equation*}

(for proof see Supp. Section~\ref{sec:theorem_1}).

This result implies that even for non-stationary wideband signals, the phase of the analytic signal in segments where the signal is almost stationary will show an almost linear increase, where the slope is an estimate of the central frequency $\bar{f}$.
This is illustrated in Figure~\ref{fig:hilbert_beamforming}e, for wideband signals with $\bar{f} = \SI{2}{\kilo\hertz}$.
As predicted, the slope of the phase function $\phi(t)$ is very well approximated by the phase progression for a sinusoid with $f_0=\SI{2}{\kilo\hertz}$.
Wideband speech samples show a similar almost linear profile of phase in the analytic signal (Figure~\ref{fig:supp_hilbert_signals}).

\subsection*{Hilbert beamforming}
Due to the linear behaviour of the phase of the analytic signal, we can take a similar beamforming approach as in the narrowband case, by constructing a weighted combination of analytic signals.
We used the central frequency estimate $\bar{f}$ to design the beamforming weights in place of the narrowband sinusoid frequency $f_0$ (see Supp.~\ref{sec:analytic_phase_beamforming}; Methods).

For narrowband sinusoidal inputs the effect of a DoA $\theta$ is to add a phase shift to each microphone input, dependent on the geometry of the array.
For a signal $a_a(t)$ of frequency $f$, we can encode this set of phase shifts with an \textit{array steering vector} as

\begin{align*}
    \textbf{s}_f(\theta) = \left( \begin{array}{c} \exp(-\textrm{j} 2 \pi f \tau_1(\theta))\\ \dots \\ \exp(-\textrm{j} 2 \pi f \tau_M(\theta)) \end{array}\right)
\end{align*}

The $M$-dimensional received analytic signal at the array is then given by $\textbf{X}_a(t) = a_a(t)\textbf{s}_f(\theta)$.
The narrowband DoA estimation problem is then solved for the chosen frequency $f$ by optimizing

\begin{align*} 
\hat{\theta} = \argmax_{\theta \in [-\pi, \pi]} \textbf{s}_f(\theta)^\herm \hat{\textbf{C}}_{x, \theta} \textbf{s}_f(\theta),
\end{align*}
where $\hat{\textbf{C}}_{x, \theta}$ is the empirical covariance of $\textbf{X}_a(t)$, which depends on both the input signal $a(t)$ and the DoA $\theta$ from which it is received (see Supp.~\ref{sec:hilbert_beamforming_weights}).

By generalising this to arbitrary wideband signals, we obtain

\begin{align*}
\hat{\textbf{C}}_{x, \theta} = 4 \int_0^\infty{\big |\bfF[a_a(t)] \big|^2 \textbf{s}_f(\theta) \textbf{s}_f(\theta)^\herm \textrm{d}f}, \label{eq:cov_analytic}
\end{align*}
where $\bfF[\cdot]$ is the Fourier transform (for proof see Supp.~\ref{sec:hilbert_beamforming_weights}).
$\hat{\textbf{C}}_{x, \theta}$ is a complex \textit{positive semi-definite} (PSD) matrix that preserves the phase difference information produced by DoA $\theta$ over all frequencies of interest $f$.

Briefly, to generate beamforming weights, we choose \textit{a priori} a desired angular precision by quantizing DoA into a grid $\clG = \{\theta_1, \dots, \theta_G\}$ with $G$ elements, with $G > M$.
We choose a representative audio signal $a(t)$, apply the Hilbert transform to obtain $a_a(t)$, and use this to compute $\hat{\textbf{C}}_{x, g}$ for $g \in G$.
The beamforming weights are obtained by finding vectors $\textbf{w}_g$ with $||\textbf{w}_g||=1$ such that $\textbf{w}_g^\herm \hat{\textbf{C}}_{x, g} \textbf{w}_g$ is maximised.
This corresponds to the singular vector with largest singular value in $\hat{\textbf{C}}_{x, g}$ and can be obtained by computing the \textit{singular value decomposition} (SVD) of $\hat{\textbf{C}}_{x, g}$.

To estimate DoA (Figure~\ref{fig:hilbert_beamforming}d) we receive the $M$-dim signal $\textbf{X}(t)$ from the microphone array and apply the Hilbert transform to obtain $\textbf{X}_a(t)$.
We then apply beamforming through the beamforming matrix $\bfW$ to obtain the $G$-dim beamformed signal $\bfX_b(t) := \bfW^\herm \bfX_a(t)$.
We accumulate the power across its $G$ components to obtain $G$-dimensional vector $\bfP = (P_{\theta_1}, \dots, P_{\theta_G})^\transp$ and estimate DoA as $\hat\theta = \argmax_{\hat\theta} P_{\hat\theta}$.

To show the benefit of our approach, we implemented Hilbert beamforming with $G = 449$ ($64 \times 7 + 1$ for an array with $M=7$ microphones with an angular oversampling of $64$) and applied it to both narrowband (Figure~\ref{fig:hilbert_beamforming}f) and wideband (\ref{fig:hilbert_beamforming}g; $B=\SI{1}{\kilo\hertz}$) signals (see Methods).
In both the best- and worst-case DoAs for the array (blue and orange curves respectively), the beam pattern (power distribution over DoA) for the wideband signal was almost identical to that for the narrowband signal, indicating that our Hilbert beamforming approach can be applied to wideband signals without first transforming them to narrowband signals.

\subsection*{Efficient online streaming implementation of Hilbert beamforming}

\begin{figure*} 
    \centering
    \includegraphics[width=120mm]{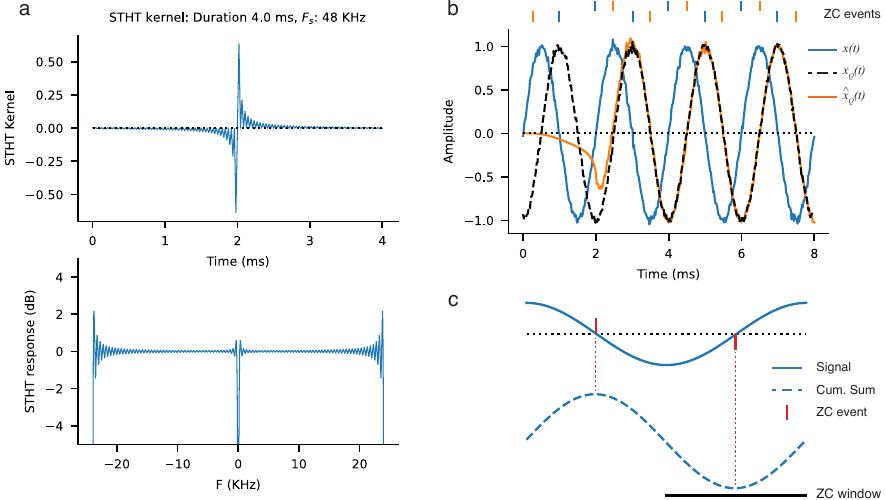}
    \caption{
        \textbf{Online Short-Time Hilbert Transform (STHT) and Robust Zero-Crossing Conjugate (RZCC) event encoding.}
        \textbf{a} The STHT kernel (top) which estimates the quadrature component $x_Q$ of a signal, and frequency response of the kernel (bottom).
        \textbf{b} A noisy narrowband input signal ($x$; blue), with the quadrature component obtained from an infinite-time Hilbert transform ($x_Q$; dashed) and the STHT-derived version ($\hat{x}_Q$; orange). Note the onset transient response of the filter before and around $t<2$ ms. Corresponding up- and down-zero crossing encoding events estimated from $x$ and $\hat{x}_Q$ are shown at top.
        \textbf{c} For a given signal (solid), zero crossings events (red) are estimated robustly by finding the peaks and troughs of the cumulative sum (dashed), within a window (ZC window).
    }
    \label{fig:STHT_RZCC}
\end{figure*}

The previous approach includes two problems that prevent an efficient streaming implementation.
Firstly, the Hilbert transform is a non-causal infinite-time operation, requiring a complete signal recording before it can be computed.
To solve this first problem, we apply an online Short-Time Hilbert Transform (STHT; Figure~\ref{fig:STHT_RZCC}a) and show that it yields a good estimate of the original Hilbert transform in the desired streaming mode.

Secondly, infinite-time power integration likewise does not lend itself to streaming operation.
The traditional solution is to average signal energy over a sliding window and update the estimate of the DoA periodically.
Our proposed method instead performs low-pass filtering in the synapses and membranes of a spiking neural network, performing the time-averaging operation natively.
 
The STHT is computed by applying a convolutional kernel $h[n]$ over a short window of length $N$ to obtain an estimation of the Quadrature component $x_Q[n]$ of a signal $x[n]$\cite{hassan_digital_2013}.
To compute the kernel $h[n]$, we made use of the linear property of the Hilbert transform (see Methods).
Briefly, the impulse response of $h[n]$ was obtained by applying the infinite-time Hilbert transform to the windowed Dirac delta signal $\delta_W[n]$ of length $W$, where $\delta_W[0]=1$ and $\delta_W[n]=0$ for $n\not = 0$ and where $W$ denotes the window length, and setting $h[n] = \textrm{Im}\left[\textbf{H}\left[\delta_W\right]\right][n]$, where $\textbf{H}$ is the Hilbert transform and $\textrm{Im}[\cdot]$ is the imaginary part of the argument.

The STHT kernel for a duration of \SI{4}{\milli\second} is shown in Figure~\ref{fig:STHT_RZCC}a (top).
The frequency response of this kernel (Figure~\ref{fig:STHT_RZCC}a, bottom) shows a predominately flat spectrum, with significant fluctuations for only low and high frequencies.
The frequency width of this fluctuation area scales proportionally to the inverse of the duration of the kernel and can be varied in case needed by changing the kernel length $W$.
In practice a bandpass filter should be applied to the audio signal before performing the STHT operation, to eliminate any distortions to low- and high-frequencies.

Figure~\ref{fig:STHT_RZCC}b illustrates the STHT applied to a noisy narrowband signal $x(t)$ (blue; in-phase component).
Following an onset transient due to the filtering settling time, the estimated STHT quadrature component $\hat{x}_Q(t)$ (orange) corresponds very closely to the infinite-time Hilbert transformed version $x_Q(t) = \textbf{H}[x(t)]$ (dashed).

\subsection*{Robust Zero-crossing spike encoding}
In order to perform the beamforming operation, we require an accurate estimation of the phase of the analytic signal.
We chose to use SNNs to implement low-power, real-time estimation of DoA; this requires an event-based encoding of the audio signals for SNN processing.
We propose a new encoding method which robustly extracts and encodes the phase of the analytic signals, ``Robust Zero-Crossing Conjugate'' (RZCC) encoding.
We estimated the zero crossings of a given signal by finding the peaks or troughs of the cumulative sum (Figure~\ref{fig:STHT_RZCC}c).
Our approach generates both up- and down-going zero-crossing events. 
Each input signal channel therefore requires 2 or 4 event channels to encode the analytic signal, depending on whether bi-polar or uni-polar zero crossing events are used.

The events produced accurately encoded the phase of the in-phase and quadrature components of a noisy signal (Figure~\ref{fig:STHT_RZCC}b; top).

Earlier works have used zero-crossing spike encoders to capture the phase of narrowband signals\cite{schoepe2023closed}.
Our approach retains the phase of the full analytic signal, and is designed to operate robustly on wideband signals.

\subsection*{SNN-based implementation of Hilbert beamforming and DoA estimation}
\begin{figure}[h!] 
    \centering
    \includegraphics[width=\linewidth]{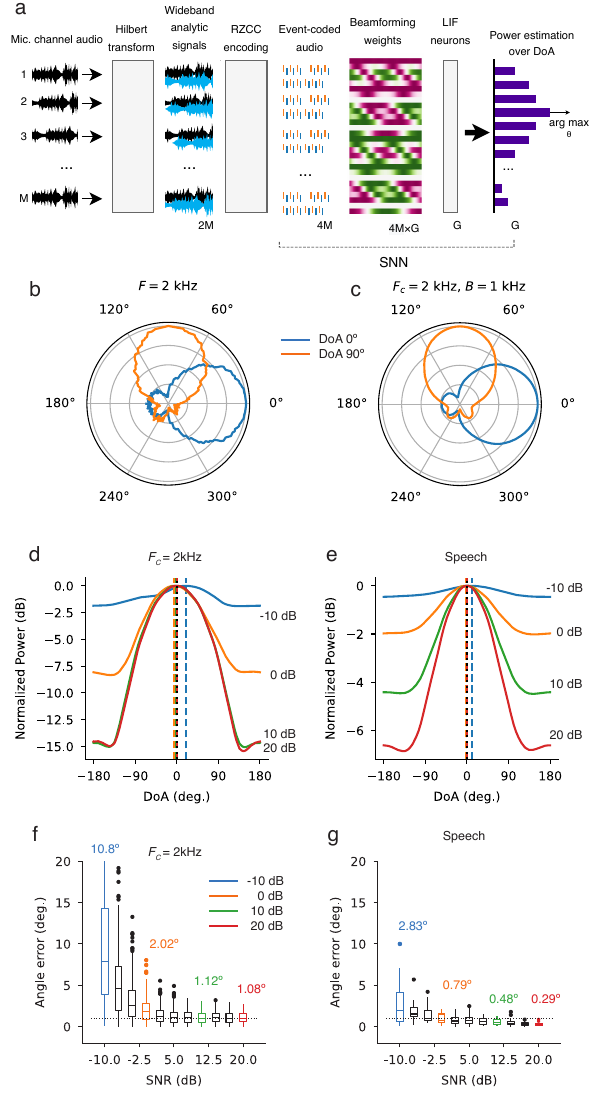}
    \caption{
        \textbf{Audio localization with STHT, RZCC encoding and DoA inference with a Spiking Neural Network (SNN).}
        \textbf{a} The pipeline for SNN implementation of our Hilbert beamforming and DoA estimation, combining Short-Time Hilbert transform; Zero-crossing conjugate encoding; analytically-derived beamforming weights; and Leaky-Integrate-and-Fire (LIF) spiking neurons for power accumulation and DoA estimation.
        \textbf{b--c} Beam patterns for SNN STHT RZCC beamforming, for narrowband (b; $F=2$ kHz) and wideband (c; $F_C=2$ kHz) signals.
        \textbf{d--e} Beam power and DoA estimates for noisy narrowband signals (d) and for noisy encoded speech (e). Dashed lines: estimated DoA.
        \textbf{f--g} DoA estimation error for noisy narrowband signals (f) and noisy speech (g). Dashed line: $1.0^{\circ}$. Annotations: Mean Absolute Error (MAE). Box plot: centre line: median; box limits: quartiles; whiskers: 1.5$\times$ inter-quartile range; points: outliers. $n=100$ random trials.
    }
    \label{fig:SNN_localization}
\end{figure}

We implemented our Hilbert beamforming and DoA estimation approach in a simulated SNN (Figure~\ref{fig:SNN_localization}).
We took the approach illustrated in Figure~\ref{fig:hilbert_beamforming}d.
We applied the STHT with kernel duration \SI{10}{\milli\second}, then used uni-polar RZCC encoding to obtain $2M$ event channels for the microphone array, for each DoA $g \in \clG$.
In place of a complex-valued analytic signal, we concatenated the in-phase and quadrature components of the $M$ signals to obtain $2M$ real-valued event channels.

Without loss of specificity, our SNN implementation used Leaky Integrate-and-Fire neurons (LIF; see Methods).
We chose synaptic and membrane time constants $\tau_s = \tau_m = \tau$, such that the equivalent low-pass filter had a \SI{3}{\deci\bel} corner frequency $f_{3\textrm{dB}}$ equal to the centre frequency of a signal of interest $f$ (i.e. $\tau = 1/ (2 \pi f)$).

We designed the SNN beamforming weights by simulating a template signal at a chosen DoA $\theta$ arriving at the array, and obtaining the resulting LIF membrane potentials $\textbf{r}_{x, \theta}(t)$, which is a $2M$ real-valued signal.
When designing beamforming weights we neglected the effect of membrane reset on the membrane potential, assuming linearity in the neuron response.
Concretely we used a chirp signal spanning \SIrange{1.5}{2.5}{\kilo\hertz} as a template.
We computed the beamforming weights by computing the sample covariance matrix
$\hat{\textbf{C}}_{x, \theta} = \textrm{E}\left[{\textbf{r}_{x, \theta}(t) \textbf{r}_{x, \theta}(t)^\transp}\right]_t$,
where $\textrm{E}[\cdot]_t$ is the expectation over time.
The beamforming weights are given by the SVD of $\hat{\textbf{C}}_{x, \theta}$. 

We reshaped the $M$-dimensioned complex signal into a $2M$-dimensioned real-valued signal, by separating real and complex components.
By doing so we therefore work with the $2M \times 2M$ real-valued covariance matrix.
Due to the phase-shifted structure relating the in-phase and quadrature components of the analytic signal, the beamforming vectors obtained by PSD from the $2M \times 2M$ covariance matrix are identical to those obtained from the $M \times M$ complex covariance matrix (See Supp. Section~\ref{app:real_complex_eq}).

Figure~\ref{fig:SNN_localization}a--b show the beam patterns for SNN Hilbert beamforming for narrowband (a) and wideband (b) signals (c.f. Figure~\ref{fig:hilbert_beamforming}f--g).
For narrowband signals, synchronisation and regularity in the event encoded input resulted in worse resolution than for wideband signals.
For noisy narrowband and noisy speech signals, beam patterns maintained good shape above \SI{0}{\decibel} SNR (Figure~\ref{fig:SNN_localization}c--d).

We implemented DoA estimation using SNN Hilbert beamforming, and measured the estimation error on noisy narrowband signals (Figure~\ref{fig:SNN_localization}e--f).
For SNR \SI{>10}{\decibel} we obtained an empirical Mean Absolute Error (MAE) of DoA estimation \SI{1.08}{\degree} on noisy narrowband signals (Figure~\ref{fig:SNN_localization}f).
DoA error increased with lower SNR, with DoA error of approx \SI{2.0}{\degree} maintained for SNRs of \SI{-1}{\decibel} and higher.
DoA error was considerably lower for noisy speech (Librispeech corpus\cite{panayotov_librispeech_2015}; Figure~\ref{fig:SNN_localization}f), with DoA errors down to \SI{0.29}{\degree}.



We deployed DoA estimation using SNN Hilbert beamforming on a seven-microphone circular array (Figure~\ref{fig:hilbert_beamforming}a) on live audio.
A sound source was placed \SI{1.5}{\meter} from the centre of the array, in a quiet office room with no particular acoustic preparation.
We collected SNN output events with the peak event rate over $G$ output channels in each \SI{400}{\milli\second} bin indicating the instantaneous DoA.
A stable DoA estimation was obtained by computing the running median over \SI{10}{\second} (see Methods).
We examined frequency bands between \SIrange{1.6}{2.6}{\kilo\hertz}, and obtained a measured MAE of \SIrange{0.25}{0.65}{\degree}.

These results are an advance on previous implementations of binaural and microphone-array source localization with SNNs.
Wall et al. implemented a biologically-inspired SNN for binaural DoA estimation, comparing against a cross-correlation approach\cite{wall2012spiking}.
They obtained an MAE of \SI{3.8}{\degree} with their SNN, and \SI{4.75}{\degree} using a cross-correlation approach, both on narrowband signals.
Escudero et al. implemented a biologically-inspired model of sound source localization using a neuromorphic audio encoding device\cite{escudero_real-time_2018}.
They obtained an MAE of \SI{2.32}{\degree} on pure narrowband signals, and \SI{3.4}{\degree} on noisy narrowband signals.
Ghosh et al. implemented a trained feed-forward SNN for binaural DoA estimation, obtaining an MAE of \SI{2.6}{\degree} on narrowband signals\cite{ghosh_spiking_2024}.
Roozbehi et al. implemented a trained recurrent SNN for binaural DoA and distance estimation, obtaining an MAE of \SI{3.4}{\degree} on wideband signals\cite{roozbehi_dynamic-structured_2024}.
Pan et al. implemented a trained recurrent SNN for DOA estimation on a circular microphone array\cite{pan2021multi}.
They reported an MAE of \SI{1.14}{\degree} under SNR \SI{20}{\decibel} traffic noise, and MAE of \SI{1.02}{\degree} in noise-free conditions\cite{pan2021multi} (see Table~\ref{tab:doa_comparison}).

We therefore set a new state-of-the-art for SNN-based DoA estimation.

\begin{table*}
	\centering
	\begin{tabular}{l l l l r l}
		\hline
		\textit{Method} & \textit{MAE$_{\SI{20}{\decibel}}$} & \textit{Filterbank size} & \textit{Power} & \textit{Compute resources} & \textit{} \\
		\textbf{Hilbert SNN} (float32) & \SI{0.29}{\degree} & --- & --- & $2 M G + 2 G$ (float32) & = \num{7184} \\
		\textbf{Hilbert SNN} (Xylo) & \SI{3.63}{\degree} & --- & \SIrange{2.53}{4.60}{\milli\watt} & $2 M G + 2 G$ (int8, int16) & = \num{7184}$^\dagger$\\
		Multi-channel RSNN  \cite{pan2021multi} & \SI{1.14}{\degree} & 40 channels & --- & $CN_r+N_r^2 + N_r G + N_r + G$ & = \num{1463712} \\
		MUSIC  \cite{schmidt1986multiple} & \SI{0.22}{\degree} & 34 channels & \SIrange{>18}{149}{\milli\watt} & $C M G$ (float32) & = \num{106862} \\
		\hline\\
	\end{tabular}
	\caption{
		\textbf{Accuracy and resource requirements for DoA estimation techniques.}
		MAE was estimated under noisy speech at \SI{20}{\decibel}.
		The Hilbert SNN and MUSIC models used $G=449$ DoA estimation channels.
		Each LIF spiking neuron was considered to have three additional parameters ($\tau_m$, $\tau_s$ and threshold $\theta$).
		The multi-channel RSNN contained $N_r=1024$ recurrent LIF neurons, and $G=360$ output LIF neurons.
		\textit{Compute resources} are the weight sizes plus state elements required for the SNN implementations, and the beamforming matrix computed in MUSIC.
		Bold text: This work.
		$^\dagger$The Xylo architecture is a mixture of 8-bit weights and 16-bit state, requiring just over a quarter of the resources of the float32 implementation. 
	}
	\label{tab:doa_comparison}
\end{table*}

\subsection*{Deployment of DoA estimation to SNN hardware}
We deployed our approach to the SNN inference architecture Xylo\cite{bos2022sub}.
Xylo implements a synchronous digital, low-bit-depth integer-logic hardware simulation of LIF spiking neurons.
The Rockpool software toolchain \cite{rockpool} includes a bit-accurate simulation of the Xylo architecture, including quantisation of parameters and training SNNs for deployment to Xylo-family hardware.
We quantised and simulated the Hilbert SNN DoA estimation approach on the Xylo architecture, using bipolar RZCC spike encoding.

Global weight and LIF parameter quantisation was performed by obtaining the maximum absolute beamforming weight value, then scaling the weights globally to ensure this maximum weight value mapped to \num{128}, then rounding to the nearest integer.
This ensured that all beamforming weights spanned the range \SIrange{-128}{128}.

\begin{figure*}[h!] 
    \centering
    \includegraphics[width=90mm]{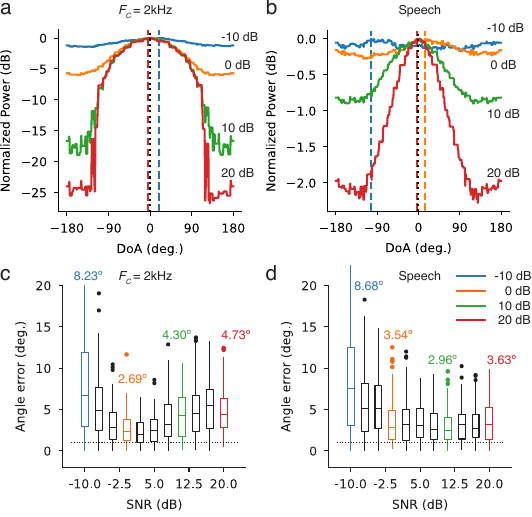}
    \caption{
        \textbf{DoA estimation using a quantised SNN architecture.}
        \textbf{a--b} Beam power and DoA estimates for noisy wideband signals (a) and noisy speech signals (b) (c.f. Figure~\ref{fig:SNN_localization}e--f).
        \textbf{c--d} DoA estimation error for noisy narrowband signals (c) and noisy speech signals (d) (c.f. Figure~\ref{fig:SNN_localization}f--g). Conventions as in Figure~\ref{fig:SNN_localization} (f--g). $n=100$ random trials.
    }
    \label{fig:xylo_snn_localization}
\end{figure*}

Spike-based DoA estimation results are shown in Figure~\ref{fig:xylo_snn_localization} and Table~\ref{tab:doa_comparison}.
For noisy speech we obtained a minimum MAE of \SI{2.96}{\degree} at SNR \SI{14}{\decibel}.



We deployed a version of Hilbert SNN DoA estimation based on unipolar RZCC encoding to the XyloAudio~2 device\cite{bos2022sub, bos2024micro}.
This is a resource-constrained low-power inference processor, supporting \num{16} input channels, up to \num{1000} spiking LIF hidden neurons, and \num{8} output channels.
We used $2\times7$ input channels for unipolar RZCC-encoded analytic signals from the $M=7$ microphone array channels, and with DoA estimation resolution of $G = 449$ as before.
We implemented beamforming by deploying the quantised beamforming weights $W$ ($14\times449$) to the hidden layer on Xylo.
We read out the spiking activity of $449$ hidden layer neurons, and chose the DoA as the neuron with highest firing rate ($\argmax$).
We measured continuous-time inference power on the Xylo processor while performing DoA estimation.

Note that the Xylo development device was not designed to provide the high master clock frequencies required for real-time operation for beamforming.
Inference results were therefore slower than real-time on the development platform.
This is not a limitation of the Xylo architecture, which could be customised to support real-time operation for beamforming and DoA estimation.
When operated at \SI{50}{\mega\hertz}, Xylo required \SI{4.03}{\second} to process \SI{2.0}{\second} of audio data, using \SI{1139}{\micro\watt} total continuous inference power.
With an efficient implementation of signed RZCC input spikes and weight sharing, the bipolar version would require an additional \SI{10}{\percent} of power consumption.
In the worst case, the bipolar RZCC version would require double the power required for the unipolar version.
Scaling our power measurements up for real-time bipolar RZCC operation, the Xylo architecture would perform continuous DoA estimation with \SIrange{2.53}{4.60}{\milli\watt} total inference power.

\subsection*{MUSIC beamforming}
\begin{figure}[hb!]
	\centering
	\includegraphics[width=\linewidth]{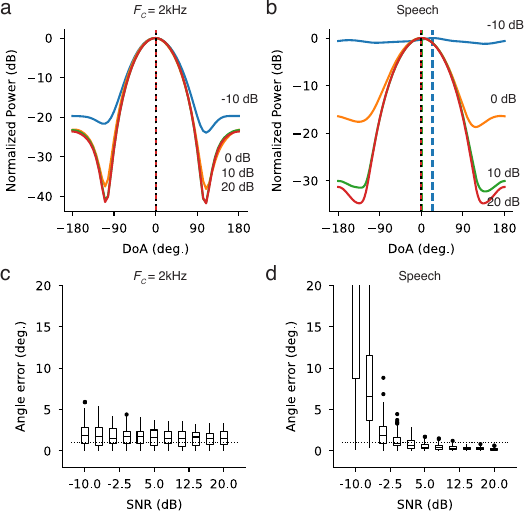}
	\caption{
	\textbf{MUSIC beamforming and DoA localization  \cite{schmidt1986multiple}.}
	\textbf{a--b} Beam patterns for noisy wideband (a) and noisy speech signals (b). Conventions as in Figure~\ref{fig:SNN_localization}.
	\textbf{c--d} DoA estimation error for noisy wideband (c) and noisy speech signals (d). Conventions as in Figure~\ref{fig:SNN_localization} (f--g). $n=100$ random trials.
	}
	\label{fig:music_localization_supp}
\end{figure}

For comparison with our method, we implemented MUSIC beamforming for DoA estimation\cite{schmidt1986multiple}, using an identical microphone array geometry.
MUSIC is a narrowband beamforming approach, following the principles in Figure~\ref{fig:hilbert_beamforming}b.
Beam patterns for MUSIC on the same circular array are shown in Figure~\ref{fig:beam_patterns_supp}.
Figure~\ref{fig:music_localization_supp} shows the accuracy distribution for MUSIC DoA estimation (same conventions as Figure~\ref{fig:SNN_localization}d).
For SNR \SI{20}{\decibel} we obtained an empirical MAE for MUSIC of \SI{1.53}{\degree} on noisy narrowband signals, and MAE of \SI{0.22}{\degree} on noisy speech.

 

To estimate the first-stage power consumption of the MUSIC beamforming method, we reviewed recent methods for low-power Fast Fourier Transform (FFT) spectrum estimation.
Recent work proposed a 128-point streaming FFT for six-channel audio in low-power \SI{65}{\nano\meter} CMOS technology, with a power supply of \SI{1.0}{\volt} and clock rate of \SI{80}{\mega\hertz} \cite{tang2019area}.
Another efficient implementation for 256-point FFT was reported for \SI{65}{\nano\meter} CMOS, using a \SI{1.2}{\volt} power supply and a clock rate of \SI{877}{\mega\hertz} \cite{hazarika2022low}.
To support a direct comparison with our method, we adjusted the reported results to align with the number of required FFT points, the CMOS technology node, the supply voltage and the master clock frequency used in Xylo (see Methods).
After appropriate scaling, we estimate these FFT methods to require \SI{149}{\milli\watt} and \SI{18.37}{\milli\watt} respectively \cite{tang2019area, hazarika2022low}.
These MUSIC power estimates include only FFT calculation and not beamforming or DoA estimation, and therefore reflect a lower bound for power consumption for MUSIC.

\subsection*{Resources required for DoA estimation}

We compared the DoA estimation errors and resources required for Hilbert SNN beamforming (this work), multi-channel RSNN-based beamforming \cite{pan2021multi} and MUSIC beamforming \cite{schmidt1986multiple} (Table~\ref{tab:doa_comparison}).
We estimated the compute resources required by each approach, by counting the memory cells used by each method.
We did not take into account differences in implementing multiply-accumulate operations, or the computational requirements for implementing FFT/DFT filtering operations in the RSNN \cite{pan2021multi} or MUSIC \cite{schmidt1986multiple} methods.

The previous state of the art for SNN beamforming and DoA estimation from a microphone array uses a recurrent SNN to perform beamforming \cite{pan2021multi}.
Their approach used a dense filterbank to obtain narrowband signal components, and a trained recurrent network with $N_r = \num{1024}$ neurons for beamforming and DoA estimation.
Their network architecture required input weights of $C N_r$ from the $C$ filterbank channels; $N_r^2$ recurrent weights for the SNN; and $N_r G$ output weights for the $G$ DoA estimation channels.
They also required $N_r + G$ neuron states for their network.
For the implementation described in their work, they required \num{1463712} memory cells for beamforming and DoA estimation.

The MUSIC beamforming approach uses a dense filterbank to obtain narrowband signals, and then weights and combines these to perform beamforming \cite{schmidt1986multiple}.
This approach requires $C M G$ memory cells.
For the implementation of MUSIC described here, \num{2514400} memory cells are required.

Since our Hilbert beamforming approach operates directly on wideband signals, we do not require a separate set of beamforming weights for each narrowband component of a source signal.
In addition, we observed that the beamforming weights for down-going RZCC input events are simply a negative version of the beamforming weights for up-going RZCC input events (see Supplementary Material \ref{sec:rzcc_supp}).
This observation suggests an efficient implementation that includes signed event encoding of the analytic signal, in the RZCC event encoding block.
This would permit a single set of beamforming weights to be reused, using the sign of the input event to effectively invert the sign of the beamforming weights.
This approach would further halve the required resources for our DoA estimation method.
Our method therefore requires $2MG$ memory cells to hold beamforming weights, and $2G$ neuron states for the LIF neurons.
For the implementation described here, our approach requires \num{7184} memory cells for beamforming and DoA estimation.

In the case of the quantised integer low-power architecture Xylo, the memory requirements are reduced further due to use of 8-bit weights and 16-bit neuron state.

Our Hilbert beamforming approach achieved very good accuracy under noisy conditions, using considerably lower compute resources than other approaches, and a fraction of the power required by traditional FFT or DFT-based beamforming methods.

\subsection*{Previous methods for beamforming and source separation using the Hilbert Transform}

Several prior works performed source localization by estimating the time delay between arrival at multiple microphones.
Kwak et al.\cite{kwak_sound_2011} applied a Hilbert transform to obtain the signal envelope, then used cross-correlation between channels to estimate the delay.
Several other works also obtained the Hilbert signal envelope and use the first peak of the amplitude envelope to estimate the delay from a sound event to a microphone in an array \cite{yang_estimation_2009, ji_hilbert_2014, barumerli_bayesian_2023, huang_research_2023}.
These are not beamforming techniques, and make no use of the phase (or full analytic signal) for source localization. 

Molla et al. applied the Hilbert transform after performing narrowband component estimation, to estimate instantaneous frequency and amplitude of incident audio \cite{molla_audio_2005}.
They then estimated inter-aural time- and level- differences (ITD and ILD), and performed standard binaural source localization based on these values, without beamforming.

Kim et al. used the Hilbert transform to obtain a version of a high-frequency input signal that can be decimated, to reduce the complexity of a subsequent FFT\cite{kim_new_2020}.
They then performed traditional frequency-domain beamforming using the narrowband signals obtained through FFT.
This is therefore distinct from our direct wideband beamforming approach.

Some works performed active beamforming on known signals, using the Hilbert transform to estimate per-channel delays prior to beamforming.\cite{abeysekera_non-linear_1999, hassan_digital_2013}
In these works, the Hilbert transform was used to obtain higher accuracy estimation of delays, where the transmitted signal is known.
We instead use the increasing phase property of the Hilbert transform to provide a general beamforming approach applicable to unknown wideband and narrowband signals.

\section*{Discussion} 

We presented a novel beamforming approach for microphone arrays, suitable for implementation in a spiking neural network (SNN).
We applied our approach to a direction-of-arrival (DoA) estimation task for far-field audio.
Our method is based on the Hilbert transform, coupled with efficient zero-crossing based encoding of the analytic signal to preserve phase information.
We showed that the phase of the analytic signal provides sufficient information to perform DoA estimation on wide-band signals, without requiring resource-intensive FFT or filterbank pre-processing to obtain narrow-band components.
We provided an efficient implementation of our method in an SNN architecture targeting low-power deployment (Xylo\cite{bos2022sub}).
By comparing our approach with state-of-the art implementations of beamforming and DoA estimation for both classical and SNN architectures, we showed that our method obtains highly accurate DoA estimation for both noisy wideband signals and noisy speech, without requiring energy- and computationally-intensive filterbank preprocessing.


Our Hilbert beamforming approach allows us to apply a unified beamforming method for both narrowband and wideband signals.
In particular, we do not need to decompose the incident signals into many narrowband components using an FFT/DFT or filterbank.
This simplifies the preprocessing and reduces resources and power consumption.

Our design for event-based zero-crossing input encoding (RZCC) suggests an architecture with signed input event channels (i.e. $+1, -1$).
The negative symmetric structure of the up-going and down-going RZCC channel Hilbert beamforming weights permits weight sharing over the signed input channels, allowing us to consider a highly resource- and power-efficient SNN architecture for deployment.

A key feature of our audio spike encoding method is to make use of the Hilbert analytic signal, making use of not only the in-phase component (i.e. the original input signal) but also the quadrature component, for spike encoding.
Previous approaches for SNN-based beamforming used a dense narrowband filterbank, obtaining almost sinusoidal signals from which the quadrature spikes are directly predicable from in-phase spikes.
Existing SNN implementations required significantly more complex network architectures for DoA estimation\cite{hu2023sound}, perhaps explained by the need to first estimate quadrature events from in-phase events, and then combine estimations across multiple frequency bands.
Our richer audio event encoding based on the STHT permits us to use a very simple network architecture for DoA estimation, and to operate directly on wideband signals.


While our approach permits unipolar RZCC event encoding and beamforming (i.e. using only up- or down-going events but not both), we observed bipolar RZCC event encoding was required to achieve high accuracy in DoA estimation.
This increases the complexity of input handling slightly, requiring 2-bit signed input events instead of 1-bit unsigned events.
The majority of SNN inference chips assume unsigned events, implying that modifications to existing hardware designs are required to deploy our method with full efficiency.


We showed that our approach achieves good DoA estimation accuracy on noisy wideband signals and noisy speech.
In practice, if a frequency band of interest is known, it may be possible to achieve even better performance by first using a wide band-pass filter to limit the input audio to a wide band of interest.

We demonstrated that engineering an SNN-based solution from first principles can achieve very high accuracy for signal processing tasks, comparable with off-the-shelf methods designed for DSPs, and state-of-the-art for SNN approaches.
Our new approach is highly appropriate for ultra-low-power SNN inference processors, and shows that SNN solutions can compete with traditional computing methods without sacrificing performance.

\begin{figure*}[h!] 
    \centering
    \includegraphics[width=\linewidth]{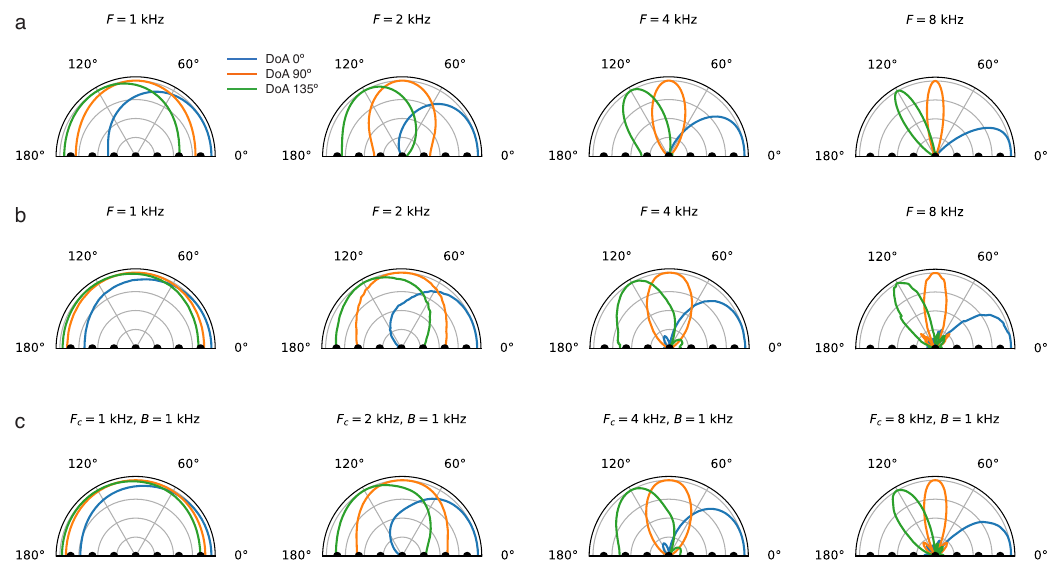}
    \caption{
        \textbf{Beam patterns for differing frequency content, for SNN Hilbert and MUSIC beamforming, on a linear microphone array.}
        We implemented a linear microphone array with seven microphones (black dots).
        \textbf{a} Beam patterns for MUSIC beamforming on narrowband signals.
        \textbf{b} Beam patterns for SNN Hilbert beamforming on narrowband signals.
        \textbf{c} Beam patterns for SNN Hilbert beamforming on wideband signals.
        Blue: DoA $\theta = \SI{0}{\degree}$ (DoA coaxial with linear array); Orange: DoA $\theta = \SI{90}{\degree}$ (orthogonal to linear array); Green: DoA $\theta = \SI{135}{\degree}$.
    }
    \label{fig:beam_patterns_linear_supp}
\end{figure*}

\begin{figure*}[h!] 
    \centering
    \includegraphics[width=\linewidth]{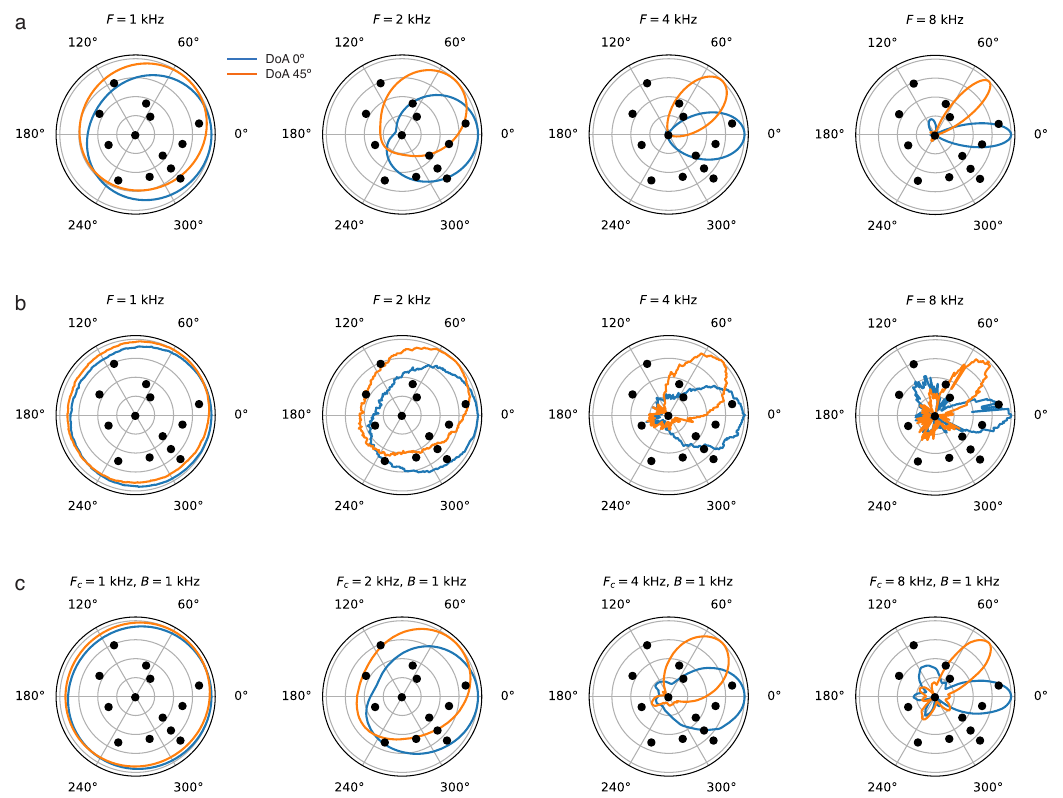}
    \caption{
        \textbf{Beam patterns for differing frequency content, for SNN Hilbert and MUSIC beamforming, on a microphone array with frozen random geometry.}
        We implemented a microphone array with 13 microphones (black dots), placed randomly within a circle of radius \SI{4.5}{\centi\meter}.
        \textbf{a} Beam patterns for MUSIC beamforming on narrowband signals.
        \textbf{b} Beam patterns for SNN Hilbert beamforming on narrowband signals.
        \textbf{c} Beam patterns for SNN Hilbert beamforming on wideband signals.
        Blue: DoA $\theta = \SI{0}{\degree}$; Orange: DoA $\theta = \SI{45}{\degree}$.
    }
    \label{fig:beam_patterns_random_supp}
\end{figure*}

Here we have demonstrated our method on a circular microphone array, but it applies equally well to alternative array geometries such as linear or random arrays with good performance (see Figures~\ref{fig:beam_patterns_linear_supp} and~\ref{fig:beam_patterns_random_supp}).

\begin{figure*}[h!] 
    \centering
    \includegraphics[width=\linewidth]{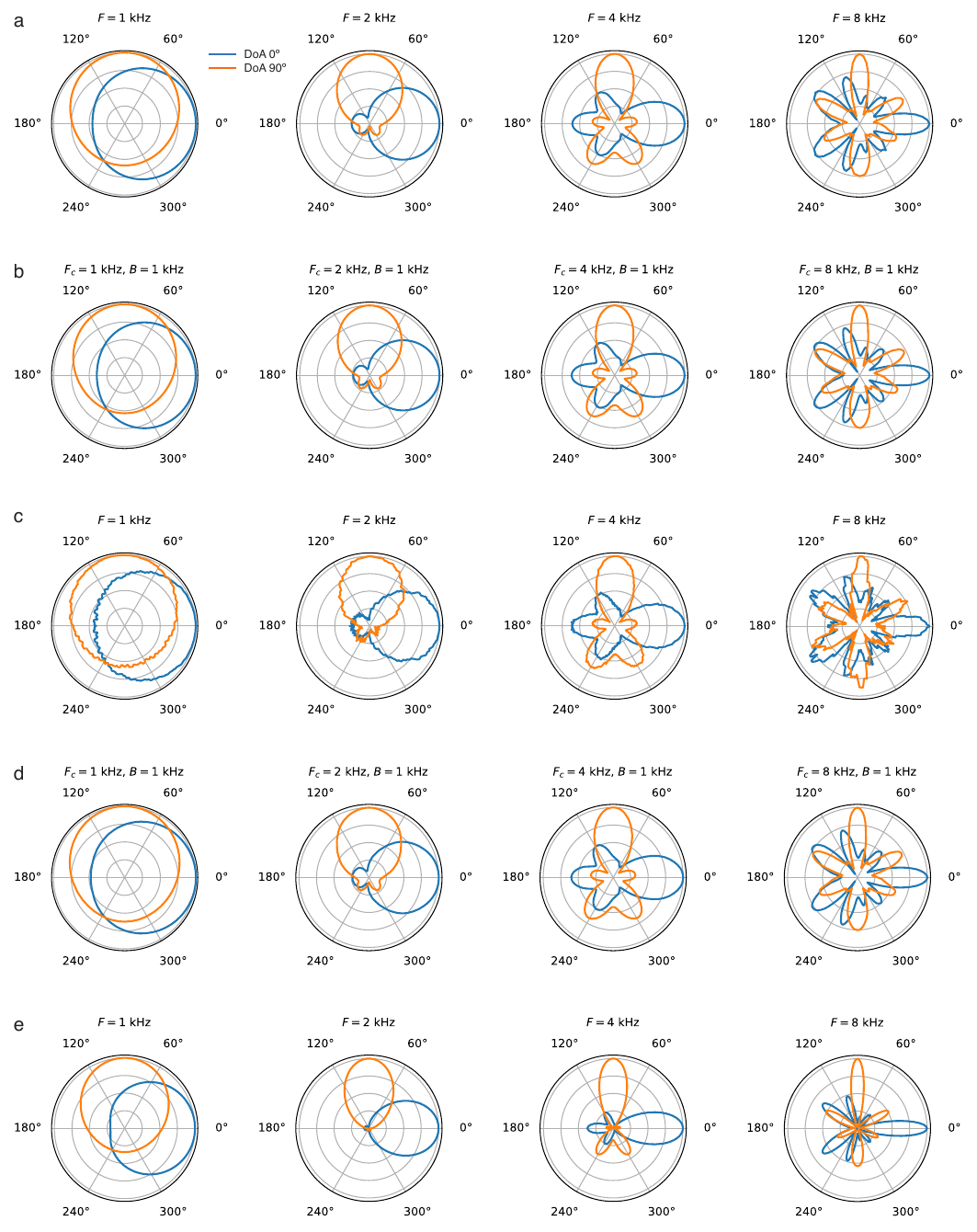}
    \caption{
        \textbf{Beam patterns for differing frequency content, for analytical and SNN implementations of Hilbert beamforming, and for MUSIC beamforming, on a circular microphone array.}
        \textbf{a} Beam patterns for Hilbert beamforming on narrowband signals.
        \textbf{b} Beam patterns for Hilbert beamforming on wideband signals.
        \textbf{c} Beam patterns for SNN Hilbert beamforming on narrowband signals.
        \textbf{d} Beam patterns for SNN Hilbert beamforming on wideband signals.
        \textbf{e} Beam Patterns for MUSIC narrowband beamforming.
        Blue: DoA $\theta = \SI{0}{\degree}$ (DoA aligned with a microphone); Orange: DoA $\theta = \SI{90}{\degree}$ (DoA midway between two microphones).
    }
    \label{fig:beam_patterns_supp}
\end{figure*}

Our novel Hilbert Transform-based beamforming method can also be applied to DSP-based signal processing solutions, by using the analytical signal without RZCC event encoding (see Figure~\ref{fig:hilbert_beamforming}f--g; Figure~\ref{fig:beam_patterns_supp}a--b).
This can improve the computational- and energy-efficiency of traditional methods, by avoiding the need for large FFT implementations.


\section*{Methods} 

\subsection*{Signal Model}\label{signal-model}
We denote the incoming signal from \(M\) microphones by \(\bfX(t)=(x_1(t), \dots, x_M(t))^\trans\) where \(x_i(t)\) denotes the time-domain signal received from the $i$-th microphone \(i\in[M]\).
We adopted a far-field scenario where the signal received from each audio source can be approximated with a planar wave with well-defined \textit{direction of arrival} (DoA).
Under this model, when an audio source at DoA \(\theta\) transmits a signal \(a(t)\), the received signal at the microphone \(i\) is given by

\[x_i(t)= \alpha a(t-\tau_i(\theta)) + z_i(t),\] where \(z_i(t)\) denotes the additive noise in microphone \(i\), where $\alpha$ is the attenuation factor (same for all microphones),  and where \(\tau_i(\theta)\) denotes the delay from the audio source to microphone \(i \in [M]\) which depends on the DoA \(\theta\).
In the far-field scenario we assumed that the attenuation parameters from the audio source to all the microphones are equal, and drop them after normalization.
We also assume that

\[\tau_i(\theta) = \tau_0 - \frac{R \cos(\theta -\theta_i)}{c},\] where
\(\tau_0 = \frac{D}{c}\) with \(D\) denoting the distance of the audio
source from the center of the array, where \(c \approx 340\) m/s is
the speed of sound in the open air, where \(R\) is the radius of the
array, and where \(\theta_i\) denotes the angle of the microphone \(i\)
in the circular array configuration as illustrated in Figure \ref{fig:hilbert_beamforming}a.

\subsection*{Wideband noise signals}
Random wideband signals were generated using coloured noise.
White noise traces were generated using iid samples from a Normal distribution: $x[n]\sim \mathcal{N}(0, 1)$, with a sampling frequency of \SI{48}{\kilo\hertz}.
These traces were then filtered using a second-order Butterworth bandpass filter from the python module \texttt{scipy.filter}.

\subsection*{Noisy speech signals}
We used speech samples from the Librispeech corpus\cite{panayotov_librispeech_2015}.
These were normalised in amplitude, and mixed with white noise to obtain a specified target SNR.

\subsection*{Beamforming for DoA Estimation}\label{sec:beamforming-for-doa-estimation}

The common approach for beamforming and DoA estimation for wideband signals is to apply DFT-like or other filterbank-based transforms to decompose the input signal into a collection of narrowband
components as 

\[\bfX(t) = \sum_ {f\in \mathcal{F}} \bfX_f(t),\]
where \(\mathcal{F} = \{f_1, f_2, \dots, f_F\}\) is a collection of
\(F = |\mathcal{F}|\) central frequencies to which the input signal is decomposed.
Beamforming is then performed on the narrowband components as follows.

When the source signal is narrowband and in the extreme case a sinusoid $x(t) = A \sin(2 \pi f_0 t)$ of frequency $f_0$, time-of-arrival to different microphones appear as a {\bf constant} phase shift

\[x_i(t) = x(t-\tau_i(\theta)) = A \sin(2 \pi f_0 t - 2\pi f_0 \tau_i(\theta)),\]
at each microphone $i\in M$ where this phase-shift depends on the DoA $\theta$. By combining the $M$ received signal $x_i(t)$ with proper weights $\bfw_\theta := (w_{1, \theta}, \dots, w_{M, \theta})^\trans$, one may zoom the array beam on a specific $\theta$ corresponding to the source DoA and obtain the beamformed signal

\begin{align*} 
	x_b(t; \theta) = \sum_{i\in[M]} w_{i,\theta} x_i(t). \label{eq:sin_bf}
\end{align*}

By performing this operation over a range of test DoAs~$\tilde\theta$, the incident DoA can be estimated by finding $\tilde\theta$ which maximises the power of the signal obtained after beamforming:

\[\argmax_{\tilde\theta} \int |x_b(t;\tilde\theta)|^2 \textrm{d}t\]

\subsection*{Hilbert beamforming}

To generate beamforming vectors, we choose a desired precision for DoA estimation by quantizing the range of DoAs into a grid \(\clG = \{\theta_1, \dots, \theta_G\}\) of size \(G\)
  where \(G = \kappa M\) where \(\kappa > 1\) denotes  the spatial oversampling
  factor.
  
We choose a template representative for the audio signal~$x(t)$, apply the HT to obtain its analytic version~$x_a(t)$, compute $\hat{\bfC}_{x, g}$ for each  angle $g \in \clG$, and design beamforming vectors $\clW_\clG:=\{\bfw_g: g\in G\}$.
We arrange the beamforming vectors as an $M\times G$ beamforming matrix $\bfW=[\bfw_0, \dots, \bfw_{G-1}]$.
In practice we choose a chirp signal for $x(t)$, spanning \SIrange{1}{3}{\kilo\hertz}.

We estimate the DoA of a target signal as follows.
We receive the \(M\)-dim time-domain signal \(\bfX(t)\) incident to the microphone array over a time-interval of duration $T$ and apply the STHT to obtain the analytic signal \(\bfX_a(t)\).

We then apply beamforming using the matrix $\bfW$ to compute the $G$-dim time-domain beamformed signal

\begin{align*}
	\bfX_b(t) = \bfW^\herm \bfX_a(t), \ t \in [0,T].
\end{align*}
We accumulate the power of the beamformed signal $\bfX_b(t)$ over the whole interval $T$, and compute the average power over the grid elements as a G-dim vector $\bfp=(p_1, \dots, p_G)^\trans$, where

\begin{align*}
	p_g &= \frac{1}{T} \int _0^T \big |[\bfX_{b}(t)]_g \big| ^2 dt  = \frac{1}{T} \int_0^T \big |\bfw_g^\herm \bfX_a(t) \big | ^2 dt.
\end{align*}

Finally we estimate the DoA of a target by identifying the DoA in $\clG$ corresponding to the maximum power, such that

\[\hat{\theta} = \argmax_{g  \in \clG} p_g.\]

\subsection*{DoA estimation with live audio}

We recorded live audio using a 7-channel circular microphone array (UMA-8 v2, radius \SI{45}{\milli\meter}), sampled at \SI{48}{\kilo\hertz}.
The microphone array and sound source (speaker width approx. \SI{6}{\milli\meter}) were placed in a quiet office environment with no particular acoustic preparation, separated by \SI{1.5}{\meter}.
We implemented Hilbert SNN beamforming, with $G = 449$ DoA bins, and a Hilbert kernel of \SI{20}{\milli\second} duration.
Chirp signals were used for template signals when generating beamforming weights, with a duration of \SI{400}{\milli\second} and a frequency range of \SIrange{1.6}{2.6}{\kilo\hertz}.
Bipolar RZCC audio to event encoding was used.
Output events were collected over the $G$ output channels, and the channel with highest event rate in each \SI{400}{\milli\second} bin was taken to indicate the instantaneous DoA estimate.
To produce a sample DoA estimation, windows of \SI{10}{\second} (25 bins) were collected, and the running median was computed over each window.
This was taken as the final DoA estimate.
Mean Absolute Error (MAE) was computed to measure the accuracy of DoA estimation.

\subsection*{LIF spiking neuron}

We used an LIF (leaky integrate and fire) spiking neuron model (See Figure \ref{fig:LIF}), in which impulse responses of synapse and neuron are given by $h_s(t) = e^{-\frac{t}{\tau_s}} u(t)$ and $h_n(t) = e^{-\frac{t}{\tau_n}} u(t)$, where $\tau_s$ and $\tau_n$ denote the time-constants of the synapse and neuron, respectively.
LIF filters can be efficiently implemented by \nth{1} order filters and bitshift circuits in digital hardware.
Here we set $\tau_s =\tau_n = \tau$.
For such a choice of parameters, the frequency response of the cascade of synapse and neuron is given by 

\begin{align*}
    |H(\omega)| = \frac{1}{1+ (\omega \tau)^2},
\end{align*}
which has a 3dB corner frequency of $f_\text{3dB} = \frac{1}{2\pi \tau}$.

\subsection*{MUSIC beamforming for DoA estimation}

We compared our proposed method with the state-of-the-art super-resolution localization method based on the MUSIC algorithm.
Our implementation of MUSIC proceeds as follows.

We apply a Fast Fourier Transform (FFT) to sequences of \SI{50}{\milli\second} audio segments, sampled at \SI{48}{\kilo\hertz} ($N = 2400$ samples), retaining only those frequency bins in the range \SIrange{1.6}{2.4}{\kilo\hertz}.
This configuration achieves an angular precision of \SI{1}{\degree} while minimising power consumption and computational complexity.
See Section \ref{sec:music_design} for details on setting parameters for MUSIC.

We computed the array response matrix for the retained FFT frequencies to target an angular precision of \SI{1}{\degree}.
To reduce the power consumption and computational complexity of MUSIC further, we retained only $F=1$ FFT frequency bin for localization, thus requiring only a single beamforming matrix.
Considering $G=225$ angular bins (where $G=225 = 7 \times 32 + 1$, for multi-mic board with $M=7$ microphones, thus, an angular oversampling factor of $32$), each array response matrix will be of dimension $7 \times 225$.

MUSIC beamforming was performed by applying FFT to frames from each microphone, choosing the highest-power frequency bin in the range \SIrange{1.6}{2.4}{\kilo\hertz}, and multiplying with the array response matrix for that freqeuncy.
We then accumulated power over time for each DoA $g \in G$, to identify the DoA with maximum power as the estimated incident DoA.

To estimate power consumption for MUSIC, we performed a survey of recent works on hardware implementation of the MUSIC algorithm, where the FFT frame size, number of microphone channels, fabrication technology, rail voltage, clock speed and resulting power consumption were all reported.
This allowed us to re-scale the reported power to match the required configuration of our MUSIC configuration, and to match the configuration of Xylo used in our own power estimates.

For example, in Ref.~\citep{tang2019area} the authors implemented a 128-point FFT in streaming mode for a 6-channel input signal in \SI{65}{\nano\meter} CMOS technology with a power supply of \SI{1.0}{\volt} and clock rate of \SI{80}{\mega\hertz}, achieving a power consumption of around \SI{10.32}{\milli\watt}.
Scaling the power to \SI{40}{\nano\meter} technology and \SI{1.1}{\volt} power supply on Xylo, assuming 7-channel input audio with a sampling rate of \SI{48}{\kilo\hertz} and FFT length $N=2048$ for our proposed MUSIC implementation, yields a power consumption of 

\begin{align*}
    P \gtrsim 10.72 \times  \frac{7}{6} \times \frac{2048}{128} \times \frac{40}{65} \times (\frac{1.1}{1.0})^2 \approx \SI{149}{\milli\watt},
\end{align*}
where the last two factors denote the effect of technology and power supply.
We made the optimistic assumption that the power consumption in streaming mode grows only proportionally to FFT frame length.
We included only power required by the initial FFT for MUSIC, and neglected the additional computational energy required for beamforming and DoA estimation itself.
Our MUSIC power estimates are therefore a conservative lower bound.

\section*{Data availability}
All data recorded as part of this study are available at \url{https://doi.org/10.5281/zenodo.13881878}\cite{haghighatshoar_muir_code_2024}.
We used speech recordings from the Librispeech corpus to measure DoA performance on human speech\cite{panayotov_librispeech_2015}.

\section*{Code availability}
All scripts to reproduce our results are available at \url{https://doi.org/10.5281/zenodo.13881878}\cite{haghighatshoar_muir_code_2024}.


\bibliographystyle{IEEEtran}
\bibliography{bibliography.bib}

\begin{thebibliography}{10}
\providecommand{\url}[1]{#1}
\csname url@samestyle\endcsname
\providecommand{\newblock}{\relax}
\providecommand{\bibinfo}[2]{#2}
\providecommand{\BIBentrySTDinterwordspacing}{\spaceskip=0pt\relax}
\providecommand{\BIBentryALTinterwordstretchfactor}{4}
\providecommand{\BIBentryALTinterwordspacing}{\spaceskip=\fontdimen2\font plus
\BIBentryALTinterwordstretchfactor\fontdimen3\font minus \fontdimen4\font\relax}
\providecommand{\BIBforeignlanguage}[2]{{%
\expandafter\ifx\csname l@#1\endcsname\relax
\typeout{** WARNING: IEEEtran.bst: No hyphenation pattern has been}%
\typeout{** loaded for the language `#1'. Using the pattern for}%
\typeout{** the default language instead.}%
\else
\language=\csname l@#1\endcsname
\fi
#2}}
\providecommand{\BIBdecl}{\relax}
\BIBdecl

\bibitem{skolnik2008radar}
M.~I. Skolnik, \emph{Radar handbook}.\hskip 1em plus 0.5em minus 0.4em\relax McGraw-Hill Education, 2008.

\bibitem{haghighatshoar2018low}
S.~Haghighatshoar and G.~Caire, ``Low-complexity massive mimo subspace estimation and tracking from low-dimensional projections,'' \emph{IEEE Transactions on Signal Processing}, vol.~66, no.~7, pp. 1832--1844, 2018.

\bibitem{li2019making}
T.~Li, L.~Fan, M.~Zhao, Y.~Liu, and D.~Katabi, ``Making the invisible visible: Action recognition through walls and occlusions,'' in \emph{Proceedings of the IEEE/CVF International Conference on Computer Vision}, 2019, pp. 872--881.

\bibitem{schmidt1986multiple}
R.~Schmidt, ``Multiple emitter location and signal parameter estimation,'' \emph{IEEE transactions on antennas and propagation}, vol.~34, no.~3, pp. 276--280, 1986.

\bibitem{roy1989esprit}
R.~Roy and T.~Kailath, ``Esprit-estimation of signal parameters via rotational invariance techniques,'' \emph{IEEE Transactions on acoustics, speech, and signal processing}, vol.~37, no.~7, pp. 984--995, 1989.

\bibitem{haykin2005cocktail}
S.~Haykin and Z.~Chen, ``The cocktail party problem,'' \emph{Neural computation}, vol.~17, no.~9, pp. 1875--1902, 2005.

\bibitem{mcdermott2009cocktail}
J.~H. McDermott, ``The cocktail party problem,'' \emph{Current Biology}, vol.~19, no.~22, pp. R1024--R1027, 2009.

\bibitem{nam2014joint}
J.~Nam, A.~Adhikary, J.-Y. Ahn, and G.~Caire, ``Joint spatial division and multiplexing: Opportunistic beamforming, user grouping and simplified downlink scheduling,'' \emph{IEEE Journal of Selected Topics in Signal Processing}, vol.~8, no.~5, pp. 876--890, 2014.

\bibitem{van1988beamforming}
B.~D. Van~Veen and K.~M. Buckley, ``Beamforming: A versatile approach to spatial filtering,'' \emph{IEEE assp magazine}, vol.~5, no.~2, pp. 4--24, 1988.

\bibitem{thompson1882li}
S.~P. Thompson, ``Li. on the function of the two ears in the perception of space,'' \emph{The London, Edinburgh, and Dublin Philosophical Magazine and Journal of Science}, vol.~13, no.~83, pp. 406--416, 1882.

\bibitem{strutt1907our}
J.~W. Strutt, ``On our perception of sound direction,'' \emph{Philosophical Magazine}, vol.~13, no.~74, pp. 214--32, 1907.

\bibitem{yin1990interaural}
T.~Yin and J.~Chan, ``Interaural time sensitivity in medial superior olive of cat,'' \emph{Journal of neurophysiology}, vol.~64, no.~2, pp. 465--488, 1990.

\bibitem{wall2012spiking}
J.~A. Wall, L.~J. McDaid, L.~P. Maguire, and T.~M. McGinnity, ``Spiking neural network model of sound localization using the interaural intensity difference,'' \emph{IEEE transactions on neural networks and learning systems}, vol.~23, no.~4, pp. 574--586, 2012.

\bibitem{escudero_real-time_2018}
\BIBentryALTinterwordspacing
E.~C. Escudero, F.~P. Pe{\~n}a, R.~P. Vicente, A.~Jimenez-Fernandez, G.~J. Moreno, and A.~Morgado-Estevez, ``Real-time neuro-inspired sound source localization and tracking architecture applied to a robotic platform,'' \emph{Neurocomputing}, vol. 283, pp. 129--139, Mar. 2018. [Online]. Available: \url{https://www.sciencedirect.com/science/article/pii/S0925231217319033}
\BIBentrySTDinterwordspacing

\bibitem{tanoni2019spiking}
G.~Tanoni, ``A spiking neural network based approach for binaural sound localization,'' 2019.

\bibitem{schoepe2023closed}
T.~Schoepe, D.~Gutierrez-Galan, J.~P. Dominguez-Morales, H.~Greatorex, A.~F. Jim{\'e}nez~Fern{\'a}ndez, A.~Linares-Barranco, and E.~Chicca, ``Closed-loop sound source localization in neuromorphic systems,'' \emph{Neuromorphic Computing and Engineering}, 2023.

\bibitem{maass1997networks}
W.~Maass, ``Networks of spiking neurons: the third generation of neural network models,'' \emph{Neural networks}, vol.~10, no.~9, pp. 1659--1671, 1997.

\bibitem{roy2019towards}
K.~Roy, A.~Jaiswal, and P.~Panda, ``Towards spike-based machine intelligence with neuromorphic computing,'' \emph{Nature}, vol. 575, no. 7784, pp. 607--617, 2019.

\bibitem{panda2020toward}
P.~Panda, S.~A. Aketi, and K.~Roy, ``Toward scalable, efficient, and accurate deep spiking neural networks with backward residual connections, stochastic softmax, and hybridization,'' \emph{Frontiers in Neuroscience}, vol.~14, p. 653, 2020.

\bibitem{cao2015spiking}
Y.~Cao, Y.~Chen, and D.~Khosla, ``Spiking deep convolutional neural networks for energy-efficient object recognition,'' \emph{International Journal of Computer Vision}, vol. 113, pp. 54--66, 2015.

\bibitem{dold2022neuro}
D.~Dold, J.~Soler~Garrido, V.~Caceres~Chian, M.~Hildebrandt, and T.~Runkler, ``Neuro-symbolic computing with spiking neural networks,'' in \emph{Proceedings of the International Conference on Neuromorphic Systems 2022}, 2022, pp. 1--4.

\bibitem{diehl2015unsupervised}
P.~U. Diehl and M.~Cook, ``Unsupervised learning of digit recognition using spike-timing-dependent plasticity,'' \emph{Frontiers in computational neuroscience}, vol.~9, p.~99, 2015.

\bibitem{akopyan2015truenorth}
F.~Akopyan, J.~Sawada, A.~Cassidy, R.~Alvarez-Icaza, J.~Arthur, P.~Merolla, N.~Imam, Y.~Nakamura, P.~Datta, G.-J. Nam \emph{et~al.}, ``Truenorth: Design and tool flow of a 65 mw 1 million neuron programmable neurosynaptic chip,'' \emph{IEEE transactions on computer-aided design of integrated circuits and systems}, vol.~34, no.~10, pp. 1537--1557, 2015.

\bibitem{davies2018loihi}
M.~Davies, N.~Srinivasa, T.-H. Lin, G.~Chinya, Y.~Cao, S.~H. Choday, G.~Dimou, P.~Joshi, N.~Imam, S.~Jain \emph{et~al.}, ``Loihi: A neuromorphic manycore processor with on-chip learning,'' \emph{Ieee Micro}, vol.~38, no.~1, pp. 82--99, 2018.

\bibitem{moraitis2020optimality}
T.~Moraitis, A.~Sebastian, and E.~Eleftheriou, ``Optimality of short-term synaptic plasticity in modelling certain dynamic environments,'' \emph{arXiv preprint arXiv:2009.06808}, 2020.

\bibitem{bos2024micro}
\BIBentryALTinterwordspacing
H.~Bos and D.~R. Muir, ``Micro-power spoken keyword spotting on {Xylo} {Audio} 2,'' Jun. 2024, arXiv:2406.15112 [cs]. [Online]. Available: \url{http://arxiv.org/abs/2406.15112}
\BIBentrySTDinterwordspacing

\bibitem{bos2022sub}
\BIBentryALTinterwordspacing
H.~Bos and D.~Muir, \emph{Sub-mW Neuromorphic SNN Audio Processing Applications with Rockpool and Xylo}.\hskip 1em plus 0.5em minus 0.4em\relax CRC Press, 2022, pp. 69--78. [Online]. Available: \url{https://ieeexplore.ieee.org/book/9967439}
\BIBentrySTDinterwordspacing

\bibitem{pan2021multi}
Z.~Pan, M.~Zhang, J.~Wu, J.~Wang, and H.~Li, ``Multi-tone phase coding of interaural time difference for sound source localization with spiking neural networks,'' \emph{IEEE/ACM Transactions on Audio, Speech, and Language Processing}, vol.~29, pp. 2656--2670, 2021.

\bibitem{hassan_digital_2013}
M.~A. Hassan and Y.~M. Kadah, ``\BIBforeignlanguage{en}{Digital {Signal} {Processing} {Methodologies} for {Conventional} {Digital} {Medical} {Ultrasound} {Imaging} {System}},'' \emph{\BIBforeignlanguage{en}{American Journal of Biomedical Engineering}}, vol.~3, no.~1, pp. 14--30, 2013, publisher: Scientific \& Academic Publishing.

\bibitem{panayotov_librispeech_2015}
\BIBentryALTinterwordspacing
V.~Panayotov, G.~Chen, D.~Povey, and S.~Khudanpur, ``Librispeech: {An} {ASR} corpus based on public domain audio books,'' in \emph{2015 {IEEE} {International} {Conference} on {Acoustics}, {Speech} and {Signal} {Processing} ({ICASSP})}, Apr. 2015, pp. 5206--5210, iSSN: 2379-190X. [Online]. Available: \url{https://ieeexplore.ieee.org/document/7178964}
\BIBentrySTDinterwordspacing

\bibitem{ghosh_spiking_2024}
\BIBentryALTinterwordspacing
M.~Ghosh, K.~G. Habashy, F.~De~Santis, T.~Fiers, D.~F. Er\c{c}elik, B.~M{\'e}sz{\'a}ros, Z.~Friedenberger, G.~B{\'e}na, M.~Hong, U.~Abubacar, R.~T. Byrne, J.~L. Riquelme, Y.~H. Liu, I.~Aizenbud, B.~A. Bicknell, V.~Bormuth, A.~Antoinetti, and D.~F.~M. Goodman, ``\BIBforeignlanguage{en}{Spiking neural network models of sound localisation via a massively collaborative process},'' 2024. [Online]. Available: \url{https://comob-project.github.io/snn-sound-localization/paper-1}
\BIBentrySTDinterwordspacing

\bibitem{roozbehi_dynamic-structured_2024}
\BIBentryALTinterwordspacing
Z.~Roozbehi, A.~Narayanan, M.~Mohaghegh, and S.-A. Saeedinia, ``Dynamic-{Structured} {Reservoir} {Spiking} {Neural} {Network} in {Sound} {Localization},'' \emph{IEEE Access}, vol.~12, pp. 24\,596--24\,608, 2024, conference Name: IEEE Access. [Online]. Available: \url{https://ieeexplore.ieee.org/document/10418107}
\BIBentrySTDinterwordspacing

\bibitem{rockpool}
D.~Muir, F.~Bauer, and P.~Weidel, ``Rockpool documentaton,'' Mar 2019.

\bibitem{tang2019area}
S.-N. Tang and Y.-H. Chen, ``Area-efficient fft kernel with improved use of gi for multistandard mimo-ofdm applications,'' \emph{Applied Sciences}, vol.~9, no.~14, p. 2877, 2019.

\bibitem{hazarika2022low}
J.~Hazarika, S.~R. Ahamed, and H.~B. Nemade, ``Low-complexity, energy-efficient fully parallel split-radix fft architecture,'' \emph{Electronics Letters}, vol.~58, no.~18, pp. 678--680, 2022.

\bibitem{kwak_sound_2011}
\BIBentryALTinterwordspacing
K.-C. Kwak, ``Sound source tracking of moving speaker using multi-channel microphones in robot environments,'' in \emph{2011 {IEEE} {International} {Conference} on {Robotics} and {Biomimetics}}, Dec. 2011, pp. 3017--3020. [Online]. Available: \url{https://ieeexplore.ieee.org/document/6181764}
\BIBentrySTDinterwordspacing

\bibitem{yang_estimation_2009}
J.~Yang, H.~Zheng, Y.~Cao, S.~Cheng, L.~Fang, and Y.~Xie, ``Estimation method for impact location of loose parts based on {Hilbert} transform,'' \emph{Journal of Mechanical Engineering}, vol.~45, no.~12, pp. 232--236, 2009.

\bibitem{ji_hilbert_2014}
T.~Ji, L.~Fang, F.~Zeng, W.~Zhang, Y.~Xie, C.~Wang, and K.~Zhang, ``Hilbert envelope loose part location method based on continuous wavelet transform,'' \emph{Atomic Energy Science and Technology}, vol.~48, no.~6, pp. 1087--1095, 2014.

\bibitem{barumerli_bayesian_2023}
\BIBentryALTinterwordspacing
R.~Barumerli, P.~Majdak, M.~Geronazzo, D.~Meijer, F.~Avanzini, and R.~Baumgartner, ``\BIBforeignlanguage{en}{A {Bayesian} model for human directional localization of broadband static sound sources},'' \emph{\BIBforeignlanguage{en}{Acta Acustica}}, vol.~7, p.~12, 2023, publisher: EDP Sciences. [Online]. Available: \url{https://acta-acustica.edpsciences.org/articles/aacus/abs/2023/01/aacus210056/aacus210056.html}
\BIBentrySTDinterwordspacing

\bibitem{huang_research_2023}
\BIBentryALTinterwordspacing
X.~Huang, R.~Xu, W.~Yu, and T.~Peng, ``Research on structural sound source localization method by neural network,'' \emph{EURASIP Journal on Advances in Signal Processing}, vol. 2023, no.~1, p.~54, May 2023. [Online]. Available: \url{https://doi.org/10.1186/s13634-023-01017-y}
\BIBentrySTDinterwordspacing

\bibitem{molla_audio_2005}
\BIBentryALTinterwordspacing
K.~Molla, K.~Hirose, and N.~Minematsu, ``Audio source separation by source localization with {Hilbert} spectrum,'' in \emph{2005 {IEEE} {International} {Symposium} on {Circuits} and {Systems}}, May 2005, pp. 5734--5737 Vol. 6, iSSN: 2158-1525. [Online]. Available: \url{https://ieeexplore.ieee.org/document/1465940}
\BIBentrySTDinterwordspacing

\bibitem{kim_new_2020}
\BIBentryALTinterwordspacing
P.~Kim, J.~H. Song, and T.-K. Song, ``\BIBforeignlanguage{en}{A new frequency domain passive acoustic mapping method using passive {Hilbert} beamforming to reduce the computational complexity of fast {Fourier} transform},'' \emph{\BIBforeignlanguage{en}{Ultrasonics}}, vol. 102, p. 106030, Mar. 2020. [Online]. Available: \url{https://linkinghub.elsevier.com/retrieve/pii/S0041624X19300587}
\BIBentrySTDinterwordspacing

\bibitem{abeysekera_non-linear_1999}
\BIBentryALTinterwordspacing
S.~S. Abeysekera, ``A non-linear algorithm for digital beamforming of a wideband linear active sonar array,'' in \emph{Nonlinear {Signal} and {Image} {Processing}}, 1999. [Online]. Available: \url{https://api.semanticscholar.org/CorpusID:39520510}
\BIBentrySTDinterwordspacing

\bibitem{hu2023sound}
F.~Hu, X.~Song, R.~He, and Y.~Yu, ``Sound source localization based on residual network and channel attention module,'' \emph{Scientific Reports}, vol.~13, no.~1, p. 5443, 2023.

\bibitem{haghighatshoar_muir_code_2024}
\BIBentryALTinterwordspacing
S.~Haghighatshoar and D.~R. Muir, ``Hilbert {SNN} code {Haghighatshoar} and {Muir} 2024,'' Oct. 2024. [Online]. Available: \url{https://doi.org/10.5281/zenodo.13881878}
\BIBentrySTDinterwordspacing

\bibitem{zygmund2002trigonometric}
A.~Zygmund, \emph{Trigonometric series}.\hskip 1em plus 0.5em minus 0.4em\relax Cambridge university press, 2002, vol.~1.

\bibitem{srm}
\BIBentryALTinterwordspacing
W.~Gerstner, ``Time structure of the activity in neural network models,'' \emph{Phys. Rev. E}, vol.~51, pp. 738--758, Jan 1995. [Online]. Available: \url{https://link.aps.org/doi/10.1103/PhysRevE.51.738}
\BIBentrySTDinterwordspacing

\bibitem{lee2023exact}
J.~H. Lee, S.~Haghighatshoar, and A.~Karbasi, ``Exact gradient computation for spiking neural networks via forward propagation,'' in \emph{International Conference on Artificial Intelligence and Statistics}.\hskip 1em plus 0.5em minus 0.4em\relax PMLR, 2023, pp. 1812--1831.

\end{thebibliography}


\beginsupplement

\clearpage
\section{Brief introduction to Hilbert transform}\label{brief-introduction-to-hilbert-transform}

The Hilbert transform (HT) is well-known in signal processing
applications. It is not, however, as widely adopted as Fourier transform family including convolution, \textit{discrete Fourier transform} (FFT/DFT), \textit{short-time Fourier transform} (STFT), etc. 

The HT is linear time-invariant (LTI) and can be described by its impulse response
\(h(t) = \frac{1}{\pi t}\) where the output $\hat{x}(t)$ for the input signal $x(t)$ is given by $\hat{x}(t) = h(t) \star x(t)$ where $\star$ denotes the convolution operation. There is some difficulty, however, due to singularity of the impulse response at
\(t=0\). By removing this singularity, one can write HT as \cite{zygmund2002trigonometric}

\begin{align}
    \hat{x}(t) = \lim_{\epsilon \to 0} \int_{\epsilon}^\infty \frac{x(t+\tau) - x(t-\tau)}{2\tau} d\tau.\label{HT_nonsingular}
\end{align}

As $h(t)$ is not zero for $t<0$ and as can be seen also from \eqref{HT_nonsingular}, HT is not a causal transform, namely, it uses both the past and future values of the signal to compute the output $\hat{x}(t)$ at each time instant $t$. 

It is always more insightful to visualize HT in the frequency domain where it can be described by
the frequency response

\[H(f) = -\textrm{j}\, \sign(f)\] where
\(\sign(.) \in \{-1, 0, +1\}\) denotes the sign function. 
%
%
For example, for a sinusoidal signal
\(x(t)=\cos(2\pi f_0 t)\) with Fourier transform

\[X(f) = \frac{1}{2} \delta(f-f_0) + \frac{1}{2} \delta(f+f_0),\] it yields

\[\hat{X}(f) = -\frac{\textrm{j}}{2} \delta(f-f_0) + \frac{\textrm{j}}{2} \delta(f+f_0),\]
which corresponds to the time-domain signal
\(\hat{x}(t) = \cos(2\pi f_0 t - \frac{\pi}{2}) = \sin(2 \pi f_0 t)\). It is seen that for sinusoidal signal, applying HT yields a simple phase shift of $-\frac{\pi}{2}$.

In this paper, we apply HT to the real-valued signal $x(t)$ received from each microphone in the array to obtain the corresponding analytic signal

\[x_a(t) := x(t) + \textrm{j} \hat{x}(t).\] 
By applying the Fourier transform to \(x_a(t) = x(t) + \textrm{j} \hat{x}(t)\),
one can see that

\begin{align*} 
	X_a(f) &= X(f) + \textrm{j} \times (-\textrm{j} \sign(f)) \times X(f)  \\
	&= (1 + \textrm{j} \times -\textrm{j} \sign(f))\times X(f) \\
	&= (1 + \sign(f))X(f) = 2 X(f) u(f),
\end{align*}
where $u(f)$ denotes the Heaviside step function.
It is seen that the spectrum of the analytic signal is zero for all negative
frequencies \(f<0\). For example, for a sinusoid \(x(t)=\cos(2\pi f_0 t)\),
one has

\[x_a(t) = \cos(2\pi f_0 t) + \textrm{j} \sin(2\pi f_0 t) = e^{\textrm{j} 2\pi f_0 t},\]
which has a Fourier transform $\delta(f-f_0)$, which is zero at negative frequencies.

\begin{remark}\label{rem:IQ}
    In signal processing literature, it is conventional to call the real-valued  signal $x(t)$ and its Hilbert transform $\hat{x}(t)$ the in-phase and quadrature components of the corresponding analytic signal $x_a(t) = x(t) + \textrm{j} \hat{x}(t)$. We use this convention throughout the paper. \hfill $\diamond$
\end{remark}

\begin{remark}
Intuitively speaking, applying HT to a real-valued signal $x(t)$ yields a complex signal $x_a(t)$, which can be illustrated as a curve in 2D complex plane. This allows to define the concept of phase and instantaneous frequency for the original signal.
For example, for a sinusoid at frequency $f_0$, the analytic signal 
\(x_a(t) = e^{\textrm{j} 2\pi f_0 t}\) has a phase $\phi(t) = 2\pi f_0 t$ which grows linearly with time $t$ and corresponds to a constant instantaneous frequency $f(t):=\frac{1}{2\pi} \frac{d\phi(t)}{dt} = f_0$. One of our main contributions is to generalize this property and show that the phase of any analytic signal  grows almost linearly in time (see, e.g., Theorem \ref{thm:phase_theorem}). This allows us to adopt conventional beamforming methods even if the signal is wideband. \hfill $\diamond$
\end{remark}

\clearpage
\section{Phase of a generic signal and its behavior}\label{extracting_phase_HT}

Let \(x(t)\) be a real-valued signal and let
\(x_a(t)= x(t) + j \hat{x}(t)\) be its corresponding analytic signal. It is conventional to define the envelope and phase of the analytic signal as

\begin{align} 
	e(t) &= \sqrt{x(t)^2 + \hat{x}(t)^2},\label{eq:env_gen}\\
	\phi(t) &= \tan^{-1}(\frac{\hat{x}(t)}{x(t)}),\label{eq:phase_gen}
\end{align} 

and write \(x_a(t)\) as
\(x_a(t) = e(t) e^{\textrm{j} \phi(t)}\).
\begin{remark}\label{rem:amp_phase}
The phase $\phi(t)$ computed from \eqref{eq:phase_gen} is in wrapped format, namely, lies in the range $[-\pi, \pi]$. So,  it may have jumps at some time instants. We need to unwrap this phase by adding or subtracting integer multiples of $2\pi$ at those jump points and glue the in-between parts together to obtain a well-defined unwrapped phase signal.
In practice, we will always work with continuous and well-behaved signals $x(t)$ for which $\hat{x}(t)$ is also continuous. For such signals, both \(e(t)\) and especially \(\phi(t)\) after unwrapping are 
continuous and well-defined functions of \(t\).  \hfill $\diamond$
\end{remark}
\begin{remark}
As a rule-of-thumb, the envelope $e(t)$ is always slowly-varying and the main variation in the analytic signal is due to the rapidly-varying phase $\phi(t)$.
For example, in the extreme case of a  sinusoid with a frequency $f_0$, the analytic signal \(x_a(t) = e^{\textrm{j} 2\pi f_0 t}\) has a constant envelope $e(t)=1$ and fast linearly-growing phase $\phi(t) = 2\pi f_0 t$. \hfill $\diamond$
\end{remark}
Let us consider the analytic signal $x_a(t)$. From inverse Fourier transform, we have that

\begin{align} 
	x_a(t) &= \int_{-\infty}^{\infty} x_a(f) e^{\textrm{j} 2\pi f t} \textrm{d}f \\
	&= \int_{0}^{\infty} 2 X(f) e^{\textrm{j} 2\pi f t} \textrm{d} f \\
	&= 2\int_{0}^{\infty} |X(f)| e^{\textrm{j} 2\pi f t + \angle X(f)} \textrm{d} f, \label{eq:an_fourier}
\end{align}
where one can see that \(x_a(t)\) is a super position of complex exponential
 \(e^{\textrm{j} 2\pi f t + \angle X(f)}\) with non-negative frequency $f$ with weights proportional to $|X(f)|$. 

As we will show, in our proposed localization method, we need the monotonic behavior of the phase $\phi(t)$ of the analytic signal $x_a(t)$. For example, for a sinusoid with frequency $f_0$, as we saw, phase $\phi(t) = 2\pi f_0 t$ grows linearly with $t$ and its slope $f_0 = \frac{1}{2\pi} \frac{d\phi(t)}{dt}$ is the parameter that specifies the spatial resolution of the DoA estimation. We would like to derive a similar counterpart for an arbitrary analytic signal $x_a(t)$ which in particular may not be narrowband.
To illustrate this, let us consider \eqref{eq:an_fourier}. Since the phase of each complex exponential component increases linearly with time $t$, we may expect that the phase of the analytic signal should be an increasing function
of \(t\).
A simple example can illustrate that this is not necessarily true. 
\begin{example}\label{ex:phase}
    Consider the signal
    
	\[x_a(t) = e_1 e^{\textrm{j} \phi_1(t)} + e_2 e^{\textrm{j} \phi_2(t)}\]
	where $0 < e_2 < e_1$ and where  both \(\phi_1(t)\) and \(\phi_2(t)\) are increasing functions of
	\(t\). Let us assume that \(e_1 > e_2\) and let us write this as

	\[x_a(t)= e_1 e^{\textrm{j} \phi_1(t)} \times (1 + \frac{e_2}{e_1} e^{\textrm{j} (\phi_2(t) - \phi_1(t)) }).\]
	Since \(\frac{e_2}{e_1} < 1\), it is not difficult to show that the phase of the second term is just  a bounded function $b(t)$ with $|b(t)| \leq \phi_{\max}$ for $\phi_{\max} \in [0, \pi]$.
	As a result, the phase of the whole signal is given by \(\phi_1(t) + b(t)\) where $\phi_1(t)$ is the phase of the stronger exponential signal.
	Since $\phi_1(t)$ is an increasing function of $t$, the whole phase $\phi(t)$ should be an \textit{almost-increasing} function of $t$.
	This is illustrated in Figure \ref{fig:two_exp_phase} for $e_1=1, e_2 = 0.9$ and for two linear phase functions $\phi_1(t)=t$ and $\phi_2(t)$. \hfill $\diamond$

	\begin{figure}[t]
    	\centering
	    \includegraphics[width=1\linewidth]{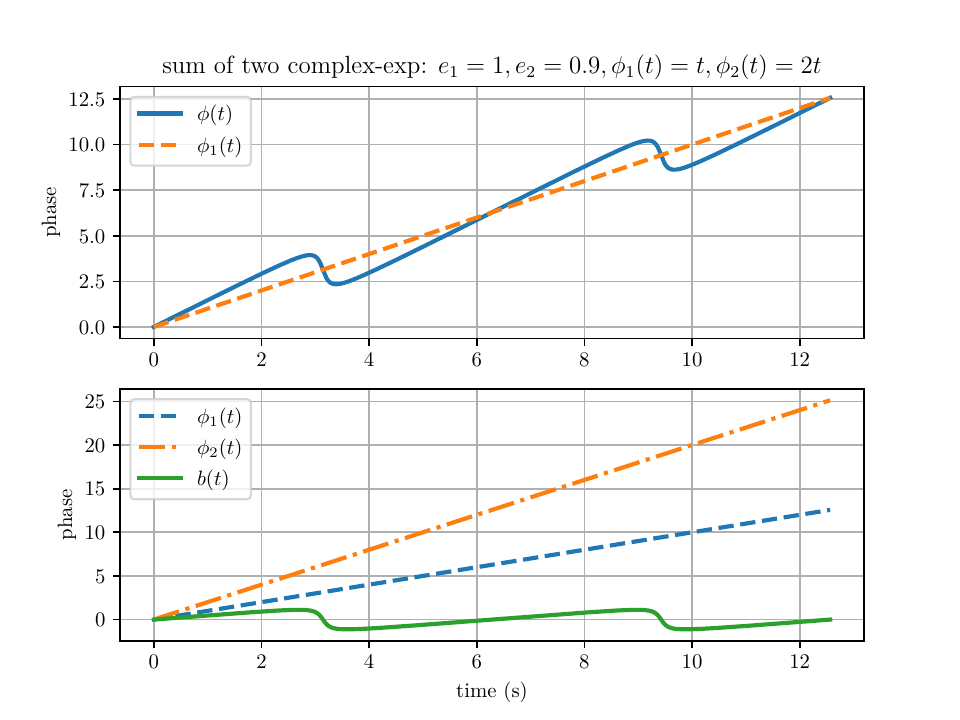}
    	\caption{Illustration of the almost-linear behavior of phase for superposition of two complex exponentials.}
	    \label{fig:two_exp_phase}
	\end{figure}
\end{example}

Example \ref{ex:phase} illustrates that in general phase may show a very complicated behavior where by  slightly modifying the amplitudes, i.e., by making \(e_1\) slightly smaller than \(e_2\), phase may switch from \(\phi_1(t)\) to \(\phi_2(t)\) (of course after neglecting the bounded additive terms). 
This extreme case of course does not happen in real-world scenarios since in practice the audio signal received from the source is a superposition of a large number of complex exponentials where none of them typically dominates the others.
``Domination'' here would imply that the amplitude of one exponential terms is larger than the sum of the amplitudes of the others.

\clearpage
\section{Almost-increasing phase property for signals with small envelope variation}\label{sec:theorem_1}
Although it is difficult to prove the almost-increasing property of the phase in the general case, we make an attempt to prove it for those signals whose envelope variation in time is quite small.
\begin{theorem}\label{thm:phase_theorem}
    Let $x_a(t)$ be the analytic signal corresponding to a real-valued signal $x(t)$ and let $e(t)$ and $\phi(t)$ be its envelope and phase respectively. Let $T>0$ be such that the time interval $t \in [0, T]$ contains major part of the energy of the signal. Define the average of $e(t)$ over this interval as $\overline{e} = \frac{1}{T} \int_0^T e(t) dt$ and suppose that $|e(t) - \overline{e}| \ll \overline{e}$ such that the variation of $e(t)$ around its average is negligible in the interval $[0, T]$. Then $\phi(t)$ is an almost-monotone function of $t$ with $\phi(T) - \phi(0) = T \overline{f}$ where $\overline{f}$  defined by
    
    \begin{align} 
    \overline{f} :=  \frac{\int _{0}^\infty f |X(f)|^2}{\int_{0}^\infty |X(f)|^2 df}\label{eq:f_bar}
    \end{align}
    is the spectral average of the frequency of the signal. \hfill $\qed$
\end{theorem}

\begin{remark}
    If we consider $p(f) = \frac{|X(f)|^2}{\int_{0}^\infty |X(f)|^2 df}$ where $\int _{0}^\infty p(f) df = 1$ as some sort of  measure that illustrates the distribution of signal energy in the frequency spectrum, we can interpret $\overline{f}$ as average of the frequency w.r.t. this measure where each frequency is weighted proportionally to the fraction of the energy it contributes to the signal. \hfill $\diamond$
\end{remark}

\begin{proof}
    We use the definition $\phi(t) = \tan^{-1}(\frac{\hat{x}(t)}{x(t)})$. Recall that as we mentioned in \Rem\,\ref{rem:amp_phase} the phase computed from this formula is in wrapped format, namely, it lies in $[-\pi, \pi]$ and may have jumps at some time instants. 
    To avoid these jumps in the proof, we take the derivative of the phase which is always local and does not need any correction or global phase unwrapping in time. We obtain
    
    \begin{align*}
        \frac{d\phi(t)}{dt} = \frac{\hat{x}(t)' x(t) - x'(t) \hat{x}(t)}{x(t)^2 + \hat{x}(t)^2}= \frac{\hat{x}(t)' x(t) - x'(t) \hat{x}(t)}{e(t)^2}.
    \end{align*}
    We then take the integral of this expression to obtain:
    
    \begin{align}
        \phi(T) &- \phi(0) = \int _{0}^T \frac{\hat{x}(t)' x(t) - x'(t) \hat{x}(t)}{e(t)^2} dt\nonumber\\
        &\stackrel{(a)}{\approx} \frac{\int _{0}^T \hat{x}(t)' x(t) - x'(t) \hat{x}(t) dt }{\overline{e}^2}\nonumber\\
        &\stackrel{(b)}{\geq } \frac{\int _{0}^T \hat{x}(t)' x(t) - x'(t) \hat{x}(t) dt }{\frac{1}{T} \int_0^T e(t)^2 dt}\nonumber\\
        &\stackrel{(c)}{\approx} T \times \frac{\int _{-\infty}^\infty \hat{x}(t)' x(t) - x'(t) \hat{x}(t) dt }{\int_{-\infty}^\infty e(t)^2 dt}\label{eq:proof_time_approx}
    \end{align}
    where
    \begin{itemize}[itemindent=0cm, leftmargin=0.4cm]
        \item in $(a)$ we used the fact that the envelop $e(t)$ changes quite slowly with time and replaced its weighting effect in the integral for sufficiently large $T$ by its average. 
        \item in $(b)$ we used the Cauchy-Schwartz inequality
        $$\overline{e}^2 \leq \overline{e^2}=: \frac{1}{T} \int_{0}^T e(t)^2 dt,$$ and that the numerator is positive as we show next.
        \item in $(c)$ we used the fact that the majority of the energy of the signal lies in the interval $[0, T]$ so we expanded the limits of the integral to $[-\infty, +\infty]$.
    \end{itemize}
    By applying the Parseval equality 
    
    \[\int _{-\infty}^\infty a(t) b(t) dt = \int_{-\infty}^{\infty} A(f)^* B(f) \textrm{d}f,\]
     for real-valued signals $a(t)$ and $b(t)$ and their Fourier transform $A(f)$ and $B(f)$ and using the fact that $\hat{X}(f) = - \textrm{j} \sign(f) X(f)$, we can show that 
     
     \begin{align*}
         \int_{-\infty}^{\infty} e(t)^2 dt &= \int_{-\infty}^{\infty} x(t)^2 + \hat{x}(t)^ 2 \textrm{d}t\\
         &= \int_{-\infty}^{\infty} |X(f)|^2 + |-\textrm{j}\sign(f) X(f)|^2 \textrm{d}f \\
         &= 2 \int_{-\infty}^{\infty} |X(f)| ^2 \textrm{d}f = 4 \int_0^\infty |X(f)|^2 \textrm{d}f.
    \end{align*}
    
    \begin{align*}
         \int_{-\infty}^{\infty} \hat{x}(t)'x(t) dt &= \int_{-\infty}^{\infty} (\textrm{j} 2\pi f \times -\textrm{j} \sign(f) X(f))^* X(f) \textrm{d}f  \\
         &= 2 \pi \int_{-\infty}^{\infty} |f| |X(f)|^2 \textrm{d}f  \\
         &= 4 \pi \int_0^\infty f |X(f)|^2 \textrm{d}f .
    \end{align*}
    
    \begin{align*}
         \int_{-\infty}^{\infty} x(t)'\hat{x}(t) dt &= \int_{-\infty}^{\infty} (\textrm{j} 2\pi f X(f))^* \times -\textrm{j} \sign(f) X(f)  \textrm{d}f \\
         &= -2 \pi \int_{-\infty}^{\infty} |f| |X(f)|^2 \textrm{d}f\\
         &= -4 \pi \int_0^\infty f |X(f)|^2 \textrm{d}f ,
     \end{align*}
    where we also used the conjugate symmetry of the Fourier transform for real-valued signals $X(-f)=X(f)^*$ which yields $|X(f)|=|X(-f)|$. Replacing these expressions in \eqref{eq:proof_time_approx} completes the proof. \hfill $\qed$
\end{proof}

\begin{example}
    One of the implications of Theorem \ref{thm:phase_theorem} is that if the signal is non-stationary with time-varying spectrum (consider, e.g., a chirp signal), the average slope of the phase of its HT in short intervals in which the signal is almost stationary will be proportional to the active signal frequency at that interval.
    Therefore, by tracking the phase of the HT one can detect the instantaneous frequency of the signal.
    Figure \ref{fig:chirp_phase} illustrates this for a chirp signal that sweeps the frequency range from $1$ to $2$ KHz with a period of 2 seconds. \hfill $\diamond$

    \begin{figure}[t]
    \centering
    \includegraphics[width=1\linewidth]{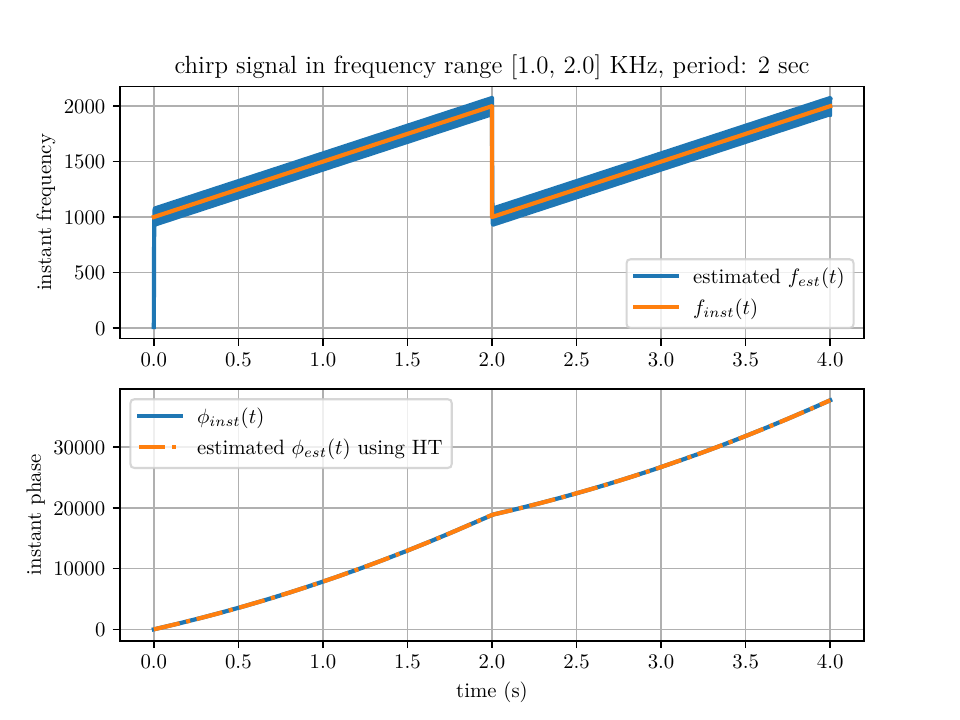}
    \caption{Illustration of tracking the instantaneous phase of a chirp signal using its HT.}
    \label{fig:chirp_phase}
    \end{figure}

\end{example}

\begin{example}
    We did some numerical simulation to investigate the precision of formula $\phi(T) - \phi(0) = T \overline{f}$. We produce white Gaussian noise and filter it with a bandpass filter with a sharp transition in the range $f \in [1, 3]$ KHz. Since the specrum of the output signal is almost flat, i.e., $|X(f)|=\text{const.}$ one can show that 
    
    \begin{align*}
        \overline{f} = \frac{\int _{f_\text{min}}^{f_\text{max}} f |X(f)|^2 \textrm{d}f }{\int _{f_\text{min}}^{f_\text{max}} |X(f)|^2 \textrm{d}f } = \frac{f_\text{min} + f_\text{max}}{2} = 2 \text{KHz}. 
    \end{align*}
    Figure \ref{fig:hilbert_beamforming}e illustrates the simulation results, where it is seen that the slope of the phase in various simulation is very close to $2$ KHz predicted by Theorem \ref{thm:phase_theorem}. \hfill $\diamond$
\end{example}

\clearpage
\section{Using phase for beamforming}\label{sec:analytic_phase_beamforming}
One of the implications of Theorem \ref{thm:phase_theorem} is that after applying the HT, the phase of the resulting analytic signal shows an almost linear growth whose slope is given by the average frequency $\overline{f}$ as in \eqref{eq:f_bar} which in the case of narrowband signals of center frequency $f_0$ yields the slope $\overline{f}=f_0$. 

As we explained before, in the case of sinusoid signals, the linear behavior of phase allows to steer the array to a specific DoA $\theta$ by a simple linear weighting of the signals received from various microphones. 
Now let us consider the audio signal \(x(t)\) and its corresponding
analytic signal \(x_a(t) = e(t) e^{j \phi(t)}\). Since both the HT and the propagation model are linear, when \(x(t)\) is
transmitted from the audio source, the analytic version of the signal
received in the microphone \(i\) is given by

\begin{align*}
    x_a(t-\tau_i(\theta)) &= e(t-\tau_i(\theta)) e^{\textrm{j} \phi(t-\tau_i(\theta))} \\
    &\approx e(t) e^{\textrm{j} \phi(t-\tau_i(\theta))}\\
    &\approx e(t) e^{\textrm{j} 2\pi \overline{f} \times (t-\tau_i(\theta))}
\end{align*}
where we used the fact that the envelope \(e(t)\) varies slowly with
time such that \(e(t) \approx e(t-\tau_i(\theta))\). It is seen that the analytic version of the received signal at microphone $i$ behaves very similarly to a sinusoid signal of frequency $\overline{f}$.

Since the phase of the HT shows a similar linear behavior, we may expect that we can steer and zoom the array on a specific DoA by applying a similar linear weighting technique with the difference that in designing the weights $\bfw_\theta= (w_{1,\theta}, \dots, w_{M,\theta})^\trans$ for a specific DoA $\theta$, we need to use the average frequency $\bar{f}$ rather than the sinusoid frequency $f_0$.

\begin{remark}
Using HT enables us to develop and apply a unified beamforming approach for both narrowband and wideband scenarios. In particular, we do not need to decompose the signal into many narrowband and almost sinusoid-like components. This simplifies the preprocessing and reduces the consumed power, which is of interest of  low-power applications we target in this paper. \hfill $\diamond$
\end{remark}


\clearpage
\section{Hilbert beamforming weights}\label{sec:hilbert_beamforming_weights}
Our main motivation for using HT and its almost-linear phase behavior is to be able to steer the array on a specific DoA by applying a simple weighting to the signals received from various microphones. In this section, we develop a step-by-step method to derive those weight parameters.

\subsection{Inspiration from Beamforming for sinusoid signals}
To gain intuition, let us start from the well-known narrowband case where $x(t) = A \cos(2\pi f t)$ and the analytic signal after applying HT is given by $x_a(t)=A e^{j 2\pi f t}$.
For such a signal, the effect of the delay in the received signal in each microphone
is to add the phase shift \(-2\pi f \tau_i(\theta)\) where $\tau_i(\theta)$ is the propagation delay from an audio source lying at DoA $\theta$ to the $i$-th microphone.  It is
convenient in array processing to write this as the \textit{array response vector} or \textit{array steering factor}

\begin{align*}
    \textbf{s}_f(\theta) = \left( \begin{array}{c} \exp(-\textrm{j} 2 \pi f \tau_1(\theta))\\ \dots \\ \exp(-\textrm{j} 2 \pi f \tau_M(\theta)) \end{array}\right)
\end{align*}
which encodes the phase variation in a narrowband signal at frequency $f$ as a function of DoA $\theta$. The $M$-dim analytic signal then can be written as $\bfX_a(t) = x_a(t) \bfs_{f}(\theta)$
For the narrowband case, one can pose DoA estimation as the following optimization problem

\begin{align*}
\hat{\theta} &= \argmax_{\theta \in [-\pi, \pi]} \int |\bfs_{f}(\theta)^\herm \bfX_a(t)|^2 \textrm{d} t \\
&= \argmax\ \bfs_{f}(\theta)^\herm \Big ( \int \bfX_a(t) \bfX_a(t)^\herm \textrm{d} t \Big ) \bfs_{f}(\theta) \\&
= \argmax\ \bfs_{f}(\theta)^\herm  \hat{\bfC}_{x, \theta}\ \bfs_{f}(\theta)
\end{align*}
where we defined the $M \times M$ empirical covariance of $\bfX_a(t)$ as 

\begin{align}
    \hat{\bfC}_{x,\theta} = \frac{1}{T} \int_0^T \bfX_a(t) \bfX_a(t)^\herm \textrm{d} t,\label{eq:hat_C}
\end{align}
where $\herm$ denotes the Hermitian transpose, where $T$ is the effective duration of the signal, and where we used the subscript $x, \theta$ to show that $\hat{\bfC}_{x, \theta}$ depends both on the input signal $x(t)$ and its DoA $\theta$. For a sinusoid with frequency $f$, since $\bfX_a(t) = A e^{j 2\pi f t} \bfs_{f}(\theta)$, one can check that $\hat{\bfC}_{x,\theta}$ is the rank-1 {\it positive semi-definite} (PSD) matrix $A^2 \bfs_{f}(\theta) \bfs_{f}(\theta)^\herm$. 
In this case, $\hat{\bfC}_{x,\theta}$ has a single non-zero singular value whose singular vector is along $\bfs_{f}(\theta)$. 
As a result, we can define the corresponding beamforming weight for a narrowband signal coming from DoA $\theta$ by computing the SVD of $\hat{\bfC}_{x, \theta}$ and using its singular vector corresponding to the largest singular value, which yields $\bfw_\theta = \bfs_f(\theta)/\sqrt{M}$. Also, once we have beamforming weights $\bfw_\theta$ for all $\theta$, we can find the DoA as 

\begin{align*}
    \hat{\theta} = \argmax_{\theta \in [-\pi, \pi]} \bfw_\theta^H \bfC_{x, \tilde{\theta}} \bfw_\theta,
\end{align*}
where $\hat{\theta}$ coincides with the DoA of the signal $\tilde{\theta}$.

\subsection{Generalization for wideband signals}
We will use the intuition gain from narrowband scenario and the linear behavior of the phase of the analytic signal to generalize beamforming for arbitrary signals $x(t)$. We first need the following theorem.

\begin{theorem}\label{thm:general_hilbert_beamforming}
    Let $x(t)$ be the real-valued input signal to the array coming from  a DoA $\theta$ and let $x_a(t)$ be its analytic version. Also let $\hat{\bfC}_{x, \theta}$ be the $M\times M$ matrix defined as in \eqref{eq:hat_C}. Then,
    
    \begin{align} 
    \hat{\bfC}_{x, \theta} = \frac{4}{T} \int _{0}^\infty |X(f)|^2 \bfs_f(\theta) \bfs_f(\theta)^\herm \textrm{d} f,\label{eq:hat_C_f}
    \end{align}
    where $X(f)$ denotes the Fourier transform of the input signal $x(t)$. \hfill $\qed$
\end{theorem}

\begin{proof}
    We write the $ij$-th component of $\hat{\bfC}_{x, \theta}$ as 
    
    \begin{align*}
        [\hat{\bfC}_{x, \theta}]_{ij} &= \frac{1}{T} \int _{-\infty}^\infty x_a(t-\tau_i(\theta)) x_a(t-\tau_j(\theta))^* \textrm{d} t\\
        &\stackrel{(a)}{=} \frac{1}{T} \int _{0}^{\infty} 2X(f) e^{-\textrm{j} 2 \pi f \tau_i(\theta)} ( 2 X(f) e^{-\textrm{j} 2\pi f \tau_j(\theta)})^* \textrm{d} f\\
        &= \frac{1}{T} \int_{0}^\infty 4 |X(f)|^2 e^{\textrm{j} 2\pi f(\tau_j(\theta) - \tau_i(\theta))} \textrm{d} f\\
        &= \frac{4}{T} \int_{0}^\infty  |X(f)|^2 [\bfs_f(\theta) \bfs_f(\theta)^\herm]_{ij} \textrm{d} f,
    \end{align*}
    where in $(a)$, we applied Parseval identity and used the fact that the Fourier transform of $x_a(t)$ is given by $2 X(f) u(f)$, thus, is zero at negative frequencies $f$.
    This completes the proof. \hfill $\qed$
\end{proof}

\begin{remark}
    It is seen from \eqref{eq:hat_C_f} that $\hat{\bfC}_{x,\theta}$ is a superposition of rank-1 PSD matrices $\bfs_f(\theta)\bfs_f(\theta)^\herm$ where the array response at frequency $f$ appears with weight proportional to the energy spectrum of the signal $|X(f)|^2$. It is also worthwhile to note that although $x(t)$ is real-valued and has both positive and negative frequency components (following conjugate symmetry), only the positive frequency components appear in $\hat{\bfC}_{x,\theta}$. This makes $\hat{\bfC}_{x,\theta}$ a complex PSD matrix that preserves the phase information due to DoA. For example, it is not difficult to check that if we had used $x(t)$ rather than $x_a(t)$ in computing $\hat{\bfC}_{x,\theta}$, we would have had $\hat{\bfC}_{x, \theta} = \hat{\bfC}_{x, \theta \pm \pi}$, thus, an ambiguity in detecting the DoA of the signal.
    \hfill $\diamond$
\end{remark}

We can then propose the following step-by-step procedure for designing the beamforming matrix:
\begin{itemize}[itemindent=0cm, leftmargin=0.4cm]
    \item choose a grid of DoAs $\clG = \{\theta_1, \dots, \theta_G\}$ where $G \gg M$ is chosen such that it yields a reasonable precision for DoA estimation.
    \item choose a candidate signal $x(t)$ and use the far-field propagation model and the array geometry to compute $\hat{\bfC}_{x,g}$ when the signal $x(t)$ is received from a generic DoA $g \in \clG$.
    \item apply the SVD to the PSD matrix $\hat{\bfC}_{x,g}$ and compute the singular vector $\bfw_g$ (normalized to have norm $1$) corresponding to the larges singular value.
    \item build the $M\times G$ beamforming matrix $\bfW$ by putting together the beamforming vectors $\bfw_g$ for all the DoA $g \in \clG$.
\end{itemize}

Once the beamforming matrix $\bfW$ was designed, we use it for DoA estimation as follows:
\begin{itemize}[itemindent=0cm, leftmargin=0.4cm]
    \item we receive the $M$-dim real-valued signal $\bfX(t)$ from $M$ microphones and apply HT to compute the $M$-dim complex analytic signal $\bfX_a(t)$.
    \item we apply beamforming to compute the $G$-dim beamformed signal $\bfX_b(t) = \bfW^H \bfX_a(t)$.
    \item we aggregate the energy of each component of $\bfX_b(t)$ over a window of duration $T$ to compute the $G$-dim power vector $\bfP = \frac{1}{T} \int_0^T |\bfX_b(t)|^2 \textrm{d}t $ where the absolute value and integration is done component-wise. 
    \item we compute the DoA of the signal by finding the grid element $g \in \clG$ with maximum power, i.e., $\hat{\theta} = \argmax_{g \in \clG} P_g$ where we used the fact that, due to multiplication with $\bfW^H$, the components of $\bfP$ are labeled with the grid elements $g \in \clG$.
\end{itemize}


\clearpage
\section{Discrete-time version of the analytic signal}
\subsection{Basic setup and extension}
In the previous sections, we intentionally decided to work with the continuous-time Hilbert transform since it allows to derive and illustrate the main ideas (e.g., behavior of phase of analytic signal and its adoption in our proposed beamforming technique) much easier. 
In practical implementations, however, we always need to work with the discrete-time sampled version of the signal.
An extension of HT  and analytic signal to the discrete-time can be easily developed. The main idea is to use the
fact the Fourier transform of the analytic signal given by
\(X_a(f) = 2 X(f) u(f)\) is zero at negative frequencies. 
This property can be used to extend HT to the discrete-time signals. Given a signal \(x[n]\) with sampling rate $f_s$ and length \(N\), we first compute its \(N\)-point DFT given by $X[k]$ where $k \in \{-\frac{N}{2}, -\frac{N}{2} + 1,  \dots, \frac{N}{2}-1\}$ when $N$ is even and $k \in \{-\frac{N-1}{2},\dots,0,\dots, \frac{N-1}{2}$ when $N$ is odd. It is known from signal processing that $X[k]$ corresponds to spectrum of the discrete-time signal $x[n]$ at the discrete frequency

\begin{align*}
    f_k = \frac{kf_s}{N}.
\end{align*}
Therefore, inspired by $X_a(f) = 2X(f)u(f)$ for the continuous-time variant, we may define the DFT of the analytic signal $x_a[n]$ by dropping the
negative frequency components in the  DFT
of $x[n]$.  There is a slight difference
between odd and even values of \(N\)\footnote{This is not an issue in the continuous-time case since $u(f)$ can be assumed to take on any value between $0$ and $1$ at the jump point $f=0$ without changing the analytic signal as variation in a single point does not change the value of the integral expression for the analytic signal. This is not, however, the case in the discrete-time case as there are finitely-many terms and changing $u[k]$ at $k=0$ changes the analytic signal.}:

\begin{itemize}[itemindent=0cm, leftmargin=0.4cm]
	\item \textbf{\(N\) is odd.} we define $u[0]=\frac{1}{2}$ at $k=0$ and set \(X_a[0] = X[0]\) and \(X_a[k] = 2 X[k]\) for \(k=1,\dots, \frac{N-1}{2}\) and \(X_a[k] = 0\) elsewhere.

	\item \textbf{\(N\) is even.} we define $u[0]=1$ and set \(x_a[k] = 2 X[k]\) for \(k=0,1, \dots, \frac{N}{2}-1\) and \(X_a[k]=0\) elsewhere.
\end{itemize}
As in the continuous-time case, by exploiting the key property that the original signal is given
by inverse DFT equation as

\[x[n] = \sum_{k} X[k] e^{j \frac{2\pi k}{N}n},\]
we may see that $x_a[n]$ consists of superposition of complex exponentials each of which has a positive discrete
frequency \(\frac{k}{N}\) (corresponding to $f_k=\frac{k f_s}{N}$ such that the phase of each term
\(e^{j \frac{2\pi k}{N}n}\) is growing linearly with \(n\).

\subsection{How to define the phase in the discrete-time case?} 
Let us consider the real-valued signal $x[n]$ and its analytic version $x_a[n]$.
As in the continuous-time case, we will call the real and imaginary part of $x_a[n]$ by in-phase and quadrature components. Similarly, we will define the envelop and phase of the analytic signal by

\begin{align*} 
	e[n] &= \sqrt{x[n]^2 + \hat{x}[n]^2},\\
	\phi[n] &= \tan^{-1}(\frac{\hat{x}[n]}{x[n]}),
\end{align*}
so that we can write

\[x_a[n] = e[n] e^{ j \phi[n]}.\]
In the case of continuous-time
signals, the phase \(\phi(t)\) of the analytic signal is a continuous
function of \(t\) with jumps of an integer multiple of $2\pi$ at only a discrete set of time instants. So there is no ambiguity in defining its unwrapped version (after removing these jumps). In the
case of discrete-time signals, consecutive samples of the phase $\phi[n]$ may have arbitrary jumps. In such a case,  we need to define the unwrapped version of the phase signal that satisfies the condition

\[|\phi[n+1] - \phi[n]| < \pi.\]
In fact, given a phase signal whose
components are all bounded in the interval \([-\pi, \pi]\) we can
convert it into its unwrapped version by modifying each term by an integer multiples of
\(2\pi\) such that the condition

\[|\phi[n+1]-\phi[n]| < \pi\]
is fulfilled. This is implemented, e.g., in \texttt{numpy.unwrap} function in Python. It is not difficult to see that, 
\(|\phi[n+1] -\phi[n]| < \pi\) is the necessary and sufficient
condition for the phase to be  uniquely recovered (of course
up to a constant shift which is an integer multiple of \(2\pi\)) from its
analytic signal \(x[n] = e[n] e^{j \phi[n]}\). In practice, this
condition can be fulfilled if the signal is sampled with a sufficiently
large sampling rate such that the variation between two consecutive phase
samples is smaller than $\pi$.

As in the continuous-time case, we can show that unwrapped phase of the 
analytic signal is an almost increasing function of $n$ since we have only complex exponentials \(e^{j \frac{2k \pi}{N} n}\)
whose phase is increasing with
\(n\) with a positive slope \(\frac{2 k \pi}{N}\). Of course, the larger the
\(k\), the faster the phase signal changes as a function of the  discrete time \(n\).

\subsection{Online approximation of Hilbert
transform for online streaming mode}\label{online-approximation-of-hilbert-transform}

%
%

One of the problems with the HT is that although it is a linear transform, it in anti-causal and requires all the samples of the signal \(x[n]\) to compute its analytic version \(x_a[n]\).
This is not typically feasible in practical implementations since it requires accumulating a large number of signal samples. To solve this issue, we develop an online streaming approximation of the HT that is applied to a windowed version of the signal of length \(W\) rather than the whole signal. 

Since HT is linear, the resulting approximate transform will should be a linear ones, thus, it  can be described by its impulse response $h[n]$.
To compute the impulse response, therefore, it is sufficient to find out how this transformation acts on a windowed Dirac delta signal $\delta_W[n]$ of length $W$ defined as $W$-dim  vector \(\bm{\delta}_W := (1, 0, \dots, 0)^\transp\).
This implies that, $h[n]$ can be easily computed by applying HT to $\bm{\delta}_W$ and setting
 
\begin{align*}
    h[n] = \Im\big[ {\bm{\delta}}_{W} \big ] _n.
\end{align*}
Now given any arbitrary real-valued signal $x[n]$ (since here the input audio is real-valued), the output of approximate HT can be written as $h[n] \star x[n]$ where $\star$ denotes the convolution operator.
We call the linear transform with impulse response $h[n]$ the \textit{Short-Time Hilbert Transform} (STHT)\cite{hassan_digital_2013}.
We define the approximate analytic signal produced by STHT by

\begin{align}
    x_a[n] = x[n-\frac{W}{2}] +  \textrm{j} h[n] \star x[n] \label{eq:stht_formula}
\end{align}
where the in-phase part is the input real-valued signal $x[n]$ whereas the quadrature part is given by $h[n] \star x[n]$.
Note that in \eqref{eq:stht_formula}, with some abuse of notation, we used the notation $\hat{x}[n]$ and $x_a[n]$ also for the output of STHT and corresponding analytic signal.
Moreover, we shifted the input signal $x[n]$ by $\frac{W}{2}$ to eliminate the delay due to the FIR filter $h[n]$ in order to align in-phase and quadrature components of STHT.

\begin{remark}
	Our proposed STHT resembles the well-known STFT: Rather than applying DFT to the whole signal samples, STFT decomposes the input signal into consecutive overlapping windows each of length $W$ and applies DFT to the signal samples within each windows.
	Similarly, in our case, by applying windowing and STHT, we convert the original signal \(x[n]\) into  its short-time (windowed) analytic version \(x_a[n]\). 
	Note that in contrast with STFT which has \(W\) filters corresponding to \(W\) frequencies at which DFT is computed, STHT has only a single FIR filter, thus, it yields a single-channel time-domain signal \(x_a[n]\) rather than the $W$-channel time-frequency-domain signal generated by STFT. \hfill $\diamond$
\end{remark}

\begin{remark}
	In the continuous-time case, HT converts the sinusoid signal \(\cos(2\pi f t)\) of frequency $f$ into \(e^{\textrm{j} 2 \pi f t}\).
	A more or less similar situation also happens in the discrete-time case with the difference for a signal of length $L$ only those sinusoids $\cos(2 \pi f/f_s n)$ whose  frequency $f \in [-\frac{f_s}{2},\frac{f_s}{2}]$ lies on the grid \(\{-\frac{f_s}{2}, -\frac{f_s}{2} + \frac{f_s}{L}, \dots, \frac{f_s}{2} - \frac{f_s}{L}, \frac{f_s}{2}\}\) with spacing \(\frac{f_s}{L}\) are converted exactly into $e^{ \textrm{j} 2\pi f/f_s n}$, where $f_s$ denotes the sampling frequency of the discrete-time signal.
	This implies that HT/STHT transformation of sinusoids will be closer to the continuous-time counterpart if the number of signal samples $L$/window length for STHT $W$ is large enough such that the frequency grid spacing \(\frac{f_s}{L}\)/$\frac{f_s}{W}$ covers the whole range $[-\frac{f_s}{2},\frac{f_s}{2}]$ very densely.  \hfill $\diamond$
\end{remark}

\subsection{Precision of STHT approximation to Hilbert transform}
Recall that we defined  HT for the discrete-time case based on DFT of the input discrete-time signal. It is well-known from signal processing that DFT as a transform decomposes any discrete-time signal $x[n]$  into a linear combination of
discrete-time sinusoids.

These two facts imply that
to evaluate the effectiveness of our proposed STHT compared with the original HT, we only need to check how effectively STHT approximates HT over the class of sinusoid signals. Moreover, since STHT is an FIR filter with impulse response $h[n]$, its response to sinusoids is again a sinusoid with the same frequency whose amplitude and phase are adjusted according to the frequency response of $h[n]$. This simply implies that we can benchmark the effectiveness STHT by comparing the frequency response of $h[n]$ with that of the original HT given by $-\textrm{j} \sign(f)$ over $f \in [-\frac{f_s}{2}, \frac{f_s}{2}]$, where $f_s$ denotes the sampling frequency of the discrete-time signal.

This is illustrated in Figure \ref{fig:STHT_RZCC}a for a STHT kernel of duration $\kerD = 4$ ms for an audio with sampling rate of $48$ KHz (thus, a window length of $W=193$ samples). It is seen that as in HT, which has a frequency response $H(f)=-\textrm{j} \sign(f)$, thus, a flat spectrum $|H(f)|=1$, STHT also has a flat response for frequencies far from the low- and high-frequency boundary $f_{\min}=\frac{1}{\kerD} = 250$ Hz and $\frac{f_s}{2}- f_{\min}$, respectively, where $\kerD$ denotes the duration of STHT kernel. The frequency width of this fluctuation can be varied in case needed by changing the kernel duration $\kerD$/kernel length $W$. In practice, however, to have a good performance, we should apply a bandpass filter to the signal before performing the STHT operation to eliminate the effect of this boundary frequency regions. 

The time-domain performance of STHT is also illustrated in Figure \ref{fig:STHT_RZCC}b for a sinusoid whose frequency lies within the flat spectrum of STHT. 
It is seen that after some transient phase of duration half kernel length
  (e.g., around $2$ ms in this example), STHT is indeed able
to recover the quadrature part of the analytic signal quite precisely.
Recall that for the sinusoid signals, quadrature part 
has a phase shift of $-\frac{\pi}{2}$ w.r.t. the original in-phase signal, as can be seen from the plots.


\clearpage
\section{Robust Zero-Crossing Conjugate (RZCC) spike encoding}\label{sec:rzcc_supp}
There is a variety of spike encoding methods that can be used to convert an analog input signal into binary spike features. For example, in applications such as keyword spotting, the spike encoding that seems to work quite well is rate-based coding where the rate of the spike in each frequency channel/filter is proportional to its instantaneous signal energy, as motivated by STFT (\textit{Short Time Fourier Transform}). This type of spike encoding, however, does not work well for localization since it is not sensitive enough to capture the difference in time of arrival of signals at various microphones, which is the crucial factor in localization. 

In this paper, instead, we use \textit{Robust Zero-Crossing Conjugate} (RZCC) spike encoding, where we call the encoding conjugate since it is applied to both in-phase and quadrature channels of the signal. In zero-crossing spike encoding, in general, a spike is produced when the output signal $x[n]$ of a filter changes sign, i.e.,

\[
s[n] = \left \{
    \begin{array}{ll}
        +1 & \text{if $x[n-1]\leq 0$ and $x[n]\geq 0$}\\
        -1 & \text{if $x[n-1]\geq 0$ and $x[n]\leq 0$}\\
        0 & \text{otherwise,}
    \end{array}
    \right.
\]
where we denoted the spike time sequence by $s[n]$ and where illustrated the up and down zero-crossings by $+1$ and $-1$, respectively.
It is not difficult to see that this type of spike encoding can be very sensitive to noise. To make it robust, we use the fact that the down/up zero-crossings of $x[n]$ correspond to the local maxima/minima of the cumulative sum signal defined by $y[n] = \sum_{-\infty}^n x[m]$. Therefore, we can increase robustness by keeping only those local maxima/minima in cumulative sum signal $y[n]$ that are maxima/minima over a window of size $\sfw$. More specifically, $n$ is announced as robust local maxima of $y[n]$ when

\begin{align*}
    y[n] > y[m] \text{ for all $m \in n-\frac{\sfw}{2}, \dots, n+\frac{\sfw}{2}$},
\end{align*}
with a similar expression holding for the local minima.
By making $\sfw$ larger, we may increase the robustness of this method to noise. The maximum spike rate produced in this type of encoding can be at most $\frac{f_s}{\sfw}$. For example, if the center frequency of the filter in the filterbank is $f_c$, it has a zero-crossing of rate $f_c$. Thus, in order not to miss the real zero-crossing, one needs to make sure that $f_c < \frac{f_s}{\sfw}$, which implies that $\sfw \leq \frac{f_s}{f_c}$. 

For example, setting $f_c=2$ KHz and $f_s=48$ KHz, we obtain $\sfw \leq 24$. So in the worst case, where the output of the filter is a zero-mean Gaussian noise, the probability that a point $n$ is announced as a robust zero-crossing is given by the probability that

\begin{align*}
    P_n:=\bP\Big [ &x[n-\frac{\sfw}{2}] <0, \dots, x[n-1] <0, \\
                    &\ \ \  x[n]>0, \dots, x[n+\frac{\sfw}{2}]>0\Big ] \\
                    &\approx \frac{1}{2^\sfw},
\end{align*}
where we used the simplifying assumption that noise samples are independent and that for a zero-mean Gaussian variable $x[m]$ we have $\bP[x[m]<0]= \frac{1}{2}$.
Thus the rate of random spikes due to noise in this method would be

\begin{align*}
r_n = P_n f_s = \frac{f_s}{2^\sfw} \text{ spike/sec.}
\end{align*}
Tab.\,\ref{tab:rzc} summarises some design choices and also quantifies the robustness of this method. For example, for an array with maximum frequency coverage of $4$ KHz, by setting $\sfw=12$, we can work with a spike rate of around $4$ K\,spike/sec when the signal is present while keeping the  random uninformative spike due to noise to as low as $12$ spike/sec when there is no audio source or when the signal received from source is very weak. It is also seen that this encoding is not that effective when the array coverage moves to higher frequency bands. For example, for a maximum frequency of $8$ KHz, we have to choose $\sfw\leq 6$ for which the noise spikes rate can be very close to signal spike rate. However, as we will see, this method is quite effective in almost all practical localization scenarios.

\begin{table}[t]
\center
\begin{tabular}{l l l l}
\hline
$\sfw$ & \textit{Filter $f_c$} & \textit{Signal spike rate $r_s$}  & \textit{Noise spike rate $r_n$} \\
6   & \SI{8}{\kilo\hertz}    & \SI{8}{ksps}           & \SI{760}{sps}            \\
8   & \SI{6}{\kilo\hertz}  & \SI{5}{ksps}           & \SI{190}{sps}            \\
10  & \SI{4.8}{\kilo\hertz}  & \SI{4.8}{ksps}          & \SI{47}{sps}              \\
12  & \SI{4}{\kilo\hertz}   & \SI{4.0}{ksps}         & \SI{12}{sps}         \\
\hline
\end{tabular}
\caption{
	Some design choices for robust zero-crossing conjugate (RZCC) spike encoding: $r_s$ and $r_n$ denote the approximate spike rate due to signal and worst-case spike rate when there is only noise.
	ksps: $\times 10^3$ spikes per second;
	sps: spikes per second.
}
\label{tab:rzc}
\end{table}

\begin{remark}
    Our RZCC spike encoding seems to be similar to neural phase spike encoding in \cite{pan2021multi} where the spikes are produced at the peak of the output of the filter $x[n]$. However, our method is completely different. First, rather than peak locations at $x[n]$, our proposed RZCC uses the zero-crossing locations of $x[n]$, which correspond to the peak locations of the cumulative sum $y[n]$ rather than that of the original signal $x[n]$. This makes a big difference since cumulative sum is a low-pass filter so it already reduces the noise power significantly and makes spike encoding quite robust to noise. Second, by detecting robust peak location of $y[n]$, we eliminate the spikes due to noise significantly. Note that this is not very important in \cite{pan2021multi} since filters are very narrowband, thus, eliminate a significant portion of the noise power but is very essential for our method since it is going to work equally well in both narrowband and wideband scenarios. \hfill $\diamond$
\end{remark}

\begin{remark}
    In general, one can see that using both up and down zero-crossings, which we call bipolar spike encoding, may indeed be redundant. We observed that this is indeed true and by careful design of beamforming vectors, one can perform SNN localization using only up zero-crossings for example, which we call unipolar spike encoding. However, this requires that one designs the weights such that the DC part of the membrane voltage signals in SNNs is eliminated to a large extent. This is not always easy, especially in the presence of weight quantization, and results in outliers in DoA estimation. These outliers, however, disappear when we use the bipolar spike encoding, as it has an interlacing of positive and negative spikes which eliminates the DC part almost perfectly. Because of this, we decided to use bipolar spike encoding, as illustrated in Figure \ref{fig:STHT_RZCC}b. \hfill $\diamond$
\end{remark}

After applying RZCC encoding for both in-phase and quadrature channels, we forward these spikes into SNN for further processing.

\subsection{SNN synapse and membrane response design}
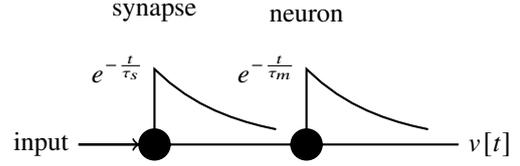
\begin{figure}
    \centering
    \begin{tikzpicture}
        \draw[->, thick] (0, 0) -- (0.8, 0); 
        \draw [thick, fill=black, ->] (0,0) node[left]{input} -- (1,0)  circle(0.2) node[above=1.5cm] {synapse} -- (3,0)  circle(0.2) node[above=1.5cm]{neuron} -- (5,0) node[right]{$v[t]$};

        \draw [thick] (1.0,0.0) 
        -- (1,1) node[left]{$e^{-\frac{t}{\tau_s}}$}
        --(1.2,0.8187307530779818)
        --(1.4,0.6703200460356393)
        --(1.6000000000000001,0.5488116360940264)
        --(1.8,0.44932896411722156)
        --(2.0,0.36787944117144233)
        --(2.2000000000000002,0.301194211912202)
        --(2.4000000000000001,0.24659696394160643)
        --(2.6,0.20189651799465538);

        \draw [thick] (3.0,0.0) 
        -- (3,1) node[left]{$e^{-\frac{t}{\tau_m}}$}
        --(3.2,0.8187307530779818)
        --(3.4,0.6703200460356393)
        --(3.6000000000000001,0.5488116360940264)
        --(3.8,0.44932896411722156)
        --(4.0,0.36787944117144233)
        --(4.2000000000000002,0.301194211912202)
        --(4.4000000000000001,0.24659696394160643)
        --(4.6,0.20189651799465538);

    \end{tikzpicture}
    \caption{Synapse and neuron impulse response in an LIF neuron.}
    \label{fig:LIF}
\end{figure}

Our method is quite generic and can be applied to a neuron with arbitrary \textit{Spike Response Model} (SRM) \cite{srm}. In this paper, as an example, we focus on LIF (leaky integrate and fire) model illustrate in Figure \ref{fig:LIF} in which impulse responses of synapse and neuron are given by $h_s(t) = e^{-\frac{t}{\tau_s}} u(t)$ and $h_n(t) = e^{-\frac{t}{\tau_n}} u(t)$ where $\tau_s$ and $\tau_n$ denote the time-constants of the synapse and neuron, respectively. Due to their exponential decay, LIF filters have the practical advantage that they can be easily implemented by \nth{1} order filters and bitshift circuits on the chip. In our design, we set $\tau_s =\tau_n = \tau$. It is not difficult to show for such a choice of parameters, the frequency response of the cascade of synapse and neuron is given by 

\begin{align*}
    |H(\omega)| = \frac{1}{1+ (\omega \tau)^2},
\end{align*}
which has $\omega_\text{3dB} = \frac{1}{\tau}$, thus, a 3dB corner frequency of $f_\text{3dB} = \frac{1}{2\pi \tau}$. In our design, we process the spike signal obtained from a filter of center frequency $f_i$ with an SNN whose parameter $\tau_i$ is selected such that $f_\text{3dB} = \frac{1}{2\pi \tau_i} =  f_i$, which implies

\begin{align*}
	\tau_i = \frac{1}{2\pi f_i}.
\end{align*}
In practice, the input signal is discrete-time and we need to approximate $h_s(t)$ and $h_n(t)$ by 

\begin{align*}
	h_s[m] &= h_s(\frac{m}{f_s})= e^{-\frac{m}{f_s\tau_s}} u[m] = \alpha_s^m u[m],\\
	h_n[m] &= h_n(\frac{m}{f_s}) = e^{-\frac{m}{f_s \tau_n}} u[m] = \alpha_n^m u[m],
\end{align*}
where $\alpha_s = e^{-\frac{1}{f_s \tau_s}}$, $\alpha_n = e^{-\frac{1}{f_s \tau_n}}$, and where $f_s$ denotes the sampling rate.
In our design, we set $\tau_s = \tau_n = \tau_i$ for filter $i$, which yields

\begin{align*} 
	\alpha_s = \alpha_n = \alpha = e^{-\frac{1}{f_s \tau}} = e^{-\frac{2\pi f_i}{f_s}}.
\end{align*} 
For example, for a filter with maximum center frequency $f_i = 3$ KHz and for a sampling frequency $f_s = 48$ KHz, we obtain that $\alpha_s=\alpha_n \approx 0.68$.

\subsection{Design of beamforming vectors}
STHT and filters in the preprocessing and also spike encoding all preserve the relative delay in signals received from various microphones. These spikes are then fed into the first layer of an SNN consisting of a linear weighting layer followed by low-pass filtering in synapse and neuron. 

Let us denote the $2M$-dim spike signal (obtained by concatenation of $M$ in-phase and $M$ quadrature spikes) by $\bfs[n]$ and let us focus on a generic neuron with weight $\bfw = (w_1, \dots, w_{2M})^\trans$. The membrane potential of this neuron at each time $m$ is given by \cite{lee2023exact}

\begin{align*}
    v[m] = \Big( \sum_{k\in[2M]} w_k s_k[m] \Big ) \star h_s[m] \star &h_n[m]\nonumber\\
    -  \theta s_o[m] \star & h_n[m] 
\end{align*}
where $\star$ denotes the convolution, where $s_o[m]$ is a discrete 0-1 valued signal denoting the output spikes, and where $-\theta s_o[m] \star h_n[m] $ denotes the refractory response due to membrane potential reset at the output spike time. 

To do beamforming, we will use the weights of the first layer in SNN. We will initialize these weights under the simplifying assumption that the neuron is linear, thus, we will neglect the refractory effect of output spikes on the membrane potential of the neuron (\textit{linear approximation}). 
Under this assumption, the membrane potential of the neuron is given by

\begin{align*}
    v[m] &= \Big( \sum_{k\in[2M]} w_k s_k[m] \Big ) \star h_s[m] \star h_n[m]\\
    &=\sum_{k\in[2M]} w_k \Big ( s_k[m]  \star h_s[m] \star h_n[m]\Big )\\
    &= \bfw^\trans \bfr_{x,\theta}[m] 
\end{align*}
where we use the fact that convolution and weighting are both linear operations, so we can exchange their order, where we defined $\bfr_{x,\theta}[m]$ as  the $2M$-dim real-valued signal whose $k$-th component is given by 

\begin{align*}
	\big[\bfr_{x,\theta}[n] \big]_k = s_k[m] \star h_s[m] \star h_n[m],
\end{align*}
and where we used the subscript $x,\theta$ to illustrate the explicit dependence of the spike signal, thus, $\bfr_{x,\theta}$, on the signal $x[n]$ and its DoA $\theta$.

Comparing with the our method for design of beamforming matrices, reveals that in the case of SNNs, we need to design beamforming matrices by applying SVD to the sample covariance matrix

\begin{align*}
    \hat{\bfC}_{x, \theta} = \frac{1}{T} \sum_{m\in[T]} \bfr_{x,\theta}[m] \bfr_{x,\theta}[m]^\trans.
\end{align*}

As we pointed out before, in contrast with $\hat{\bfC}_{x,\theta}$ which was an $M\times M$ complex PSD matrix, $\hat{\bfC}_{x, \theta}$ in the SNN version is a real-valued PSD matrix but of a larger dimension $2M \times 2M$. 
One can indeed show that, for a generic signal, the singular vectors are incompatible between the real and complex version. However, as we prove in Section \ref{app:real_complex_eq}, this does not happen when the real and imaginary components are the in-phase and quadrature parts of an underlying analytic signal, where in that case there is a one-to-one correspondence between the singular vectors (due to correlation between the real and imaginary parts).

\clearpage
\section{Equivalence of real- and complex-valued beamformer design}\label{app:real_complex_eq}
Let as denote the  $M$-dim analytic signal by $\bfX_a(t) = \bfX_i(t) + \textrm{j} \bfX_q(t)$. We first prove the following lemma.

\begin{lemma}
    Let $\hat{\bfC}_a = \bfA+ j\bfB$ be the sample covariance of the analytic signal $\bfX_a(t)$. Then the sample covariance $\hat{\bfC}_r$ of the real-valued $2M$-dim signal $\bfX_r(t) := (\bfX_i(t)^\trans, \bfX_q(t)^\trans)^\trans$ is given by 
    
    \begin{align*}
        \hat{\bfC}_r = \frac{1}{2}
        \left [ 
            \begin{matrix}
                \bfA & -\bfB \\ \bfB & \bfA
            \end{matrix}
        \right].
    \end{align*}
    Therefore, there is a one-to-one correspondence between $\hat{\bfC}_a$ and $\hat{\bfC}_r$. \hfill $\qed$
\end{lemma}

\begin{remark}
    As we will show in the proof of this lemma, this correspondence between $\hat{\bfC}_a$ and $\hat{\bfC}_r$ holds only because $\bfX_i(t)$ and $\bfX_q(t)$ are I/Q conjugates and does not hold for arbitrary signals $\bfX_i(t)$ and $\bfX_q(t)$. \hfill $\diamond$
\end{remark}

\begin{proof}
Let us define $\hat{\bfC}_{\alpha\beta}$ with $\alpha, \beta \in \{i, q\}$ the matrix

\begin{align*}
    \hat{\bfC}_{\alpha \beta} = \frac{1}{T} \int_0 ^T \bfX_\alpha(t) \bfX_\beta(t)^\trans dt.
\end{align*}
By applying the Parseval's identity and using the fact that in the Fourier domain we have $\bfX_q(f) = -\textrm{j} \sign(f) \bfX_i(f)$, it is straightforward to show that $\hat{\bfC}_{\alpha \alpha} = \hat{\bfC}_{\beta \beta}$ and $\hat{\bfC}_{\alpha \beta} = - \hat{\bfC}_{\beta \alpha}$. Therefore, by defining $\bfA:= 2\hat{\bfC}_{ii} = 2\hat{\bfC}_{qq}$ and $\bfB := 2\hat{\bfC}_{qi} = -2\hat{\bfC}_{iq}$, one can see that the sample covariance of $\bfX_r(t)$ is given by

\begin{align*}
        \hat{\bfC}_r = \frac{1}{2}
        \left [ 
            \begin{matrix}
                \bfA & -\bfB \\ \bfB & \bfA
            \end{matrix}
        \right].
\end{align*}
Moreover, one can check that the sample covariance of the complex analytic signal $\bfX_a(t) = \bfX_i(t) + \textrm{j} \bfX_q(t)$ can be written as 

\begin{align*}
    \hat{\bfC}_a &= \hat{\bfC}_{ii} + \hat{\bfC}_{qq} + \textrm{j} \hat{\bfC}_{qi} - \textrm{j} \hat{\bfC}_{iq} \\
    &= 2 \hat{\bfC}_{ii} + \textrm{j} 2\hat{\bfC}_{qi}\\
    &= \bfA + \textrm{j} \bfB.
\end{align*}
This completes the proof.
\end{proof}

The next theorem makes the final connection between the real-valued and complex-valued  covariance matrices.

\begin{theorem}
    Let $\hat{\bfC}_a$ and $\hat{\bfC}_r$ be the order $M$ complex and order $2M$ real covariance matrices. Let us denote the singular values and singular vectors of $\hat{\bfC}_a$ by $\lambda_1 \geq \dots \geq \lambda_M$ and $\bfw_1, \dots, \bfw_M$ where $\bfw_i = \bfu_i + \textrm{j} \bfv_i$ where $\bfu_i, \bfv_i \in \bR^M$ denote the real and complex part of the singular vector $\bfw_i$, $i\in[M]$. Then $\hat{\bfC}_r$ has singular values $\lambda_1, \lambda_1, \dots, \lambda_M, \lambda_M$ where each $\lambda_i$ has a multiplicity of $2$ (thus, a complete set of $2M$ singular values) and corresponds to two real-valued singular vectors 
    
    \begin{align*}
        \bfw_i = \bfu_i + \textrm{j} \bfv_i \in \bC^M \to \left [\begin{matrix} \bfu_i \\ \bfv_i \end{matrix}\right ], \left [\begin{matrix} -\bfv_i \\ \bfu_i \end{matrix}\right ] \in \bR^{2M}.
    \end{align*}
\end{theorem}

\begin{proof}
The proof simply follows by writing the singular vector definition for $\hat{\bfC}_a = \bfA+ \textrm{j} \bfB$. More specifically, denoting a generic singular value by $\lambda$ and corresponding singular vector by $\bfw = \bfu+ \textrm{j} \bfv$, we have that 

\begin{align*}
    \hat{\bfC}_a \bfw = \lambda \bfw \to  (\bfA+ \textrm{j} \bfB) \times (\bfu + \textrm{j} \bfv) = \lambda (\bfu + \textrm{j} \bfv)
\end{align*}
yield the following set of equations

\begin{align*}
    &\bfA \bfu - \bfB \bfv = \lambda \bfu \\
    &\bfB \bfu + \bfA \bfv = \lambda \bfv.
\end{align*}
This immediately implies that $\left [ \begin{matrix} \bfu \\ \bfv \end{matrix} \right ] \in \bR^{2M}$ is the real-valued singular vector corresponding to the singular value $\lambda$. By changing the order of the equations, one can also verify that $\left [ \begin{matrix} -\bfv \\ \bfu \end{matrix} \right ] \in \bR^{2M}$ corresponds to the singular vector for the same singular value $\lambda$. This completes the proof.
\end{proof}

\begin{remark}\label{rem:c_hat_rotation}
    Since singular vectors $\left [ \begin{matrix} \bfu \\ \bfv \end{matrix} \right ]$ and $\left [ \begin{matrix} -\bfv \\ \bfu \end{matrix} \right ]$ are orthogonal and correspond to the same singular value, one can verify that  any vector of the form 
    
    \begin{align*}
        \left [ \begin{matrix} \cos(\gamma) \bfu - \sin(\gamma) \bfv \\ \cos(\gamma) \bfv  + \sin(\gamma) \bfu\end{matrix} \right ],
    \end{align*}
    for an arbitrary $\gamma \in [0, 2\pi]$ is also a singular vector of $\hat{\bfC}_r$ with the same singular value $\lambda$. This indeed implies that the singular subspace corresponding to $\lambda$ is 2-dim. Seen in the complex domain, this result simply implies that if $\bfw = \bfu + j \bfv$ is a singular vector of the complex covariance matrix $\hat{\bfC}_a$ so is its rotated version $e^{j\gamma} \bfw$ for an arbitrary $\gamma \in [0, 2\pi]$. \hfill $\diamond$
\end{remark}

\begin{remark}
    One of the important implications of \Rem\ref{rem:c_hat_rotation} is that while  evaluating the overlap/correlation between two different DoAs $\theta, \theta'$, we need to $(i)$ compute the real-valued singular vectors $\left [ \begin{matrix} \bfu_\theta \\ \bfv_\theta \end{matrix} \right ]$ and $\left [ \begin{matrix} \bfu_{\theta'} \\ \bfv_{\theta'} \end{matrix} \right ]$ $(ii)$ construct the complex valued counterpart $\bfw_\theta = \bfu_\theta + j\bfv_\theta$ and $\bfw_{\theta'} = \bfu_{\theta'} + j \bfv_{\theta'}$, and then $(iii)$ the complex-valued inner product $|\inp{\bfw_\theta}{\bfw_{\theta'}}|$ to obtain the beam pattern as function of $\theta'$ when the array points at a specific DoA $\theta$. If we denote the 2-dim linear subspace generated by $\bfu_\theta, \bfv_\theta$ and $\bfu_{\theta'}, \bfv_{\theta'}$ by $L_2(\theta)$ and $L_2(\theta')$, we can write the beam pattern $b_{\theta, \theta'}$ as 
    
    \begin{align*}
        b_{\theta, \theta'} = \argmax_{\bm{\nu} \in L_2(\theta), \bm{\mu} \in L_2(\theta')} \inp{\bm{\nu}}{\bm{\mu}},
    \end{align*}
    which implies that in computing the beam pattern, one should take into account the maximum overlap/interference between two DoAs to target the least selectivity of the array. This makes sense intuitively and is illustrated much easier in the complex counterpart. \hfill $\diamond$
\end{remark}

\clearpage
\section{MUSIC beamforming and localisation}\label{sec:music_design}
In the wideband applications addressed in this paper, one has significant flexibility in choosing signal bandwidth, number and frequency subbands, number of samples in each FFT frame, etc. for the implementation of MUSIC algorithm.
To obtain the minimum power consumption for MUSIC, we adopt the following design setup:
\begin{itemize}[itemindent=0cm, leftmargin=0.3cm]
\item applying FFT to a sequence consisting of $N$ audio samples yields its frequency content equally sampled at $N$ points in the frequency range $[-\frac{f_s}{2}, \frac{f_s}{2}]$ Hz where $f_s = 48$\,KHz is the sampling rate of the input audio in multi-mic board we are using.  Since we are mainly working in the frequency band $1\sim 3$ KHz for audio (and also with real-valued audio signal), we will need only those FFT frequency bins that lie in the range $1\sim 3$ KHz.

\item each FFT bin $i$ registers signal frequencies in the range $[(i-1)\frac{f_s}{N}, (i+1)\frac{f_s}{N}]$. So $N$, thus, FFT resolution should be designed to be high enough so that narrowband beamforming can be realized. Otherwise, due to possible shift in frequency, the array response vector, thus, the beam pattern may be shifted in the angular domain,  reducing the precision of DoA estimation. 

\item assuming a slice of $\Delta t = \frac{N}{f_s}$ of input audio consisting of $N$ samples on which the FFT is applied, the frequency resolution is $\Delta f = \frac{1}{\Delta t}=\frac{f_s}{N}$. So the worst frequency shift happens when we move from $f \to f+\Delta f$ when $f$ is the lowest frequency, say, $1$ KHz. Here we should make sure that narrow-band approximation holds, i.e., $\frac{f + \Delta f}{f} = 1 + \frac{\Delta f}{f} \approx 1$, say with a precision of $1$\% or more to make sure that we can achieve an angular precision of $1^\circ$. 

\item setting $f=1$ KHz, this implies that we need at least a resolution of $\Delta f = 10$ Hz, thus, at least an FFT frame of duration $100$ ms. If we are more inclined towards higher frequencies to obtain better angular resolution, we can set $f=2$ KHz and target a frequency resolution of $\Delta f = 20$ Hz, thus, an FFT frame of duration of $50$ ms.
\item with the audio sampling rate $48$ KHz in multi-mic board, this requires at least $N = 2400 \sim 4800$ signal samples within the FFT frame to fulfill narrowband approximation in MUSIC at high/low frequency. 

\item we set the FFT frame length to be a power of $2$ for more efficient implementation and to be even more in favor of the methods in the literature, we use the smallest frame size $N=2048$ to target the lowest power consumption for MUSIC.

\end{itemize}

To achive an angular precision of \SI{1}{\degree}, we use $G=225$ angular bins (where $G=225 = 7 \times 32 + 1$, for multi-mic board with $M=7$ microphones, thus, an angular oversampling factor of $32$). As a result, the array response matrix at each frequency (used as beamforming matrix at that frequency) will be of dimension $7 \times 225$.

In the most naive case, beamforming at each frequency requires multiplying  the $M \times G$ array response matrix with the $M$-dim signal obtained from FFT. This is assuming that we only use a $F=1$ FFT frequency bin for localization. If we have more frequency bins $F > 1$, the computational complexity, thus, power consumption grows proportionally to $F$. Again to favor MUSIC algorithm, we assume that we are using only $F=1$ FFT frequency bin for localization.

In general, matrix multiplication at the beamforming stage can indeed be more computationally demanding than computing the FFT itself. For special array geometries, however, one may be able to use array structure to avoid direct multiplication. For example, in linear arrays, multiplication with the $M\times G$ array response matrix is equivalent to doing another FFT in the angle domain where the FFT size (considering angular bins) may be chosen to be a power of 2 for efficient implementation. This, however, seems not to be easy in the circular array geometry used in this paper. To favor MUSIC algorithm, we focus on power consumption for FFT implementation. This will definitely yield a lower bound on the power consumption of MUSIC.

\onecolumn
\clearpage
\section{Beam patterns for the several beamforming methods on a circular array}

\onecolumn
\clearpage
\section{Beam patterns for the several beamforming methods on a linear array}

\onecolumn
\clearpage
\section{Beam patterns for the several beamforming methods on a random array}

\onecolumn
\clearpage
\section{Direction of Arrival estimation on a quantised SNN inference architecture}



\end{document}